\documentclass[journal]{IEEEtran}
\IEEEoverridecommandlockouts

\usepackage[utf8]{inputenc} 
\usepackage[T1]{fontenc}
\usepackage{url}              % provides \url{...}
\usepackage{hyperref}
\usepackage[cmex10]{amsmath}  % Use the [cmex10] option to ensure complicance
\usepackage{amssymb,amsfonts, amsthm}
\usepackage{dsfont}
\usepackage{bbm}
\usepackage{mleftright}       % fix to wrong spacing of \left-,
\mleftright                   % \middle- \right-commands 

\usepackage{graphicx}         % provides \includegraphics{...} to
\usepackage{booktabs}         % fixes poor spacing in tables and
\usepackage[dvipsnames]{xcolor}
\usepackage{tikz}
\usetikzlibrary{arrows}
\usetikzlibrary{positioning}
\usepackage{pgf}
\usepackage[backend=biber,url=false,style=ieee,dashed=false,isbn=false,doi=false,clearlang=false]{biblatex}
\bibliography{2nd-order-idf}

\DeclareSourcemap{
  \maps[datatype=bibtex]{
    \map[overwrite]{
      \step[fieldsource=shortjournal, fieldtarget=journaltitle]
    }
  }
}

\usepackage{pgfplots}
\pgfplotsset{compat=1.18}

\definecolor{plot0}{RGB}{36,107,143}  % IDf blue
\definecolor{plot1}{RGB}{47,125,79}   % IDF green
\definecolor{plot2}{RGB}{106,76,147}  % timid/bold purple

\usepackage{xcolor}
\usepackage{marginnote}

\newcommand{\indic}{\mathbbm{1}}
\newcommand*{\expect}[2][]{\ensuremath{\mathbb{E}_{#1}\left[#2\right]}}

\newcommand{\Var}{\mathrm{Var}}
\renewcommand{\vec}[1]{\mathbf{#1}}
\newcommand{\hb}{H_{\mathrm{b}}}

\newcommand{\na}{n_1}
\newcommand{\nb}{n_2}
\newcommand{\ntot}{n}
\renewcommand\vec{\mathbf}

\theoremstyle{definition}
\newtheorem{definition}{Definition}
\newtheorem{remark}{Remark}
\newtheorem{example}{Example}

\theoremstyle{plain}
\newtheorem{theorem}{Theorem}
\newtheorem{lemma}{Lemma}
\newtheorem{corollary}{Corollary}

\begin{document}

\title{The Second-Order Coding Rate of \\ Identification with Noiseless Feedback} 

\author{\IEEEauthorblockN{Marcel~A.~Mross,~\IEEEmembership{Graduate~Student~Member,~IEEE}, Christian~Deppe, \IEEEmembership{Senior~Member,~IEEE}, and~Eduard~A.~Jorswieck,~\IEEEmembership{Fellow,~IEEE}}
\thanks{Parts of this work were presented at IEEE ITW 2025 \cite{mross_second-order_2025}. The authors are with the Institute for Communications Technology, Technische Universität Braunschweig, Germany (email: \{m.mross, christian.deppe, e.jorswieck\}@tu-braunschweig.de). This work was partly supported by the Federal Ministry of Research, Technology, and Space (BMFTR) of Germany in the programs of "Souverän. Digital. Vernetzt." Joint project xG-RIC, project identification number: 16KIS2435 (M. Mross, E. Jorswieck) and 6G-life², project identification number: 16KIS2415 (C. Deppe). The work of C. Deppe was also partly supported by the BMFTR project Q-STARS, project identification number 16KIS2602.
}}

\maketitle

\begin{abstract}
We investigate second-order coding rates for identification over discrete memoryless channels (DMCs) with noiseless feedback. For deterministic encoding, we derive an achievable second-order rate and prove its optimality for the class of simple-dispersion channels. For randomized encoding, we characterize the second-order capacity for arbitrary DMCs.  Achievability follows from a two-phase construction that generates common randomness and transmits a hash, using threshold-based decoding and a tag-transmission phase of logarithmic length. The converse bounds combine counting arguments with martingale methods to handle the dependence introduced by feedback. For weakly symmetric channels, the deterministic identification-feedback dispersion coincides with the transmission dispersion, while the randomized identification-feedback dispersion vanishes when the entropy-maximizing output distribution is uniform.
\end{abstract}


\section{Introduction}\label{sec:introduction}
\IEEEPARstart{I}{dentification} via channels is a communication paradigm that was proposed by JáJá \cite{jaja_identification_1985} and Ahlswede and Dueck \cite{ahlswede_identification_1989} as an alternative to the classical message transmission problem. In a two-terminal communication system, both the transmitter and the receiver have an associated message, $i$ and $i'$, respectively, which is unknown to the other party. The task of the receiver is to find out whether $i=i'$, and the transmitter's task is to encode $i$ and transmit it to the receiver over a noisy channel in such a way that the total number $N$ of such messages that can be reliably identified is maximized. Thus, the receiver task reduces to a binary hypothesis problem, while in Shannon's message transmission paradigm, the receiver has to estimate the exact message that was sent. The seminal result from \cite{ahlswede_identification_1989} shows that randomization at the encoder can lead to an impressive scaling behavior: while for message transmission, the total number of messages scales as $M = 2^{nC}$, the number of identification messages scales double exponentially as $2^{2^{nC}}$, where $n$ denotes the blocklength and $C$ denotes the channel capacity in both cases. Also without randomization at the encoder (deterministic identification), higher rates than in message transmission can be achieved, even though they are not on the double exponential scale \cite{ahlswede_identification_1999, salariseddigh_deterministic_2022, colomer_optimal_2026-1}. 

If the transmitter at time instance $t$ has access to the received symbol $Y_{t-1}$ from the previous time instance, we speak of noiseless feedback. A classic result by Shannon \cite{shannon_zero_1956} shows that feedback cannot increase the capacity of point-to-point message transmission over a discrete memoryless channel (DMC). However, for identification, feedback can lead to significant improvements in performance: the capacity is characterized by the output entropy of the channel, meaning that more noise in the channel can even improve the identification capacity as long as the transmission capacity is still strictly positive \cite{ahlswede_identification_1989-1}. Feedback also helps to let the message set size $N$ of deterministic identification scale double exponentially with $n$. This is due to the fact that the remarkable performance of identification codes is enabled by common randomness between transmitter and receiver, and through the feedback link, the channel noise can be leveraged to generate such common randomness \cite{ahlswede_identification_1989-1}. This also leads to the result that for Gaussian channels, the identification-feedback capacity is infinite  \cite{labidi_identification_2021}. 
Identification techniques have been extended to various multi-user scenarios \cite{oohama_converse_2013-1, bracher_identification_2017, rosenberger_deterministic_2023}, secrecy-constrained settings \cite{ahlswede_new_1995, labidi_secure_2020, rosenberger_capacity_2023}, and explicit code constructions have been actively explored \cite{verdu_explicit_1993, derebeyoglu_performance_2020, gunlu_code_2022, von_lengerke_codes_2026}. The identification problem can be utilized in emerging applications in goal-oriented communications \cite{getu_survey_2024}, 6G \cite{cabrera_6g_2022} and molecular communication \cite{salariseddigh_deterministic_2023}. We refer the reader to \cite{cabrera_6g_2022, von_lengerke_codes_2026, labidi_identification_2026} for in-depth explanations and surveys on the topic of identification via channels.

The results on identification mentioned so far are capacity results, meaning that they are valid in the limit $n \to \infty$. Especially due to the double exponential scaling, it is to be expected that the number $N$ of reliably identifiable messages is already very large at very small blocklengths $n$. Older works already considered error exponents to get insights into the finite-blocklength behavior \cite{ahlswede_identification_1989}. Since then, the analysis of \emph{second-order coding rates}\footnote{In older literature on identification, the term \emph{second-order rate} is used to describe the double logarithmic definition of rates that is used to capture the double exponential scaling behavior. Our usage of this term (see Def. \ref{def:idf-det-second-order}) is not to be confused with that.} has gained some prominence to describe the behavior of rates in the so-called \textit{Central Limit Theorem (CLT)-regime}, where the bound on the error probability is fixed and the convergence of the rates towards the capacity is analyzed as $n$ approaches infinity. The analysis of second-order rates for transmission over the DMC was pioneered by Strassen \cite{strassen_asymptotische_1962} and significant progress was made by the seminal works of Hayashi \cite{hayashi_information_2009} and Polyanskiy, Poor, and Verdú \cite{polyanskiy_channel_2010}. Since then, second-order rates have been derived for several multi-user scenarios \cite{tan_dispersions_2014, molavianjazi_second-order_2015, yang_wiretap_2019,  mross_gaussian_2024}, transmission with feedback \cite{polyanskiy_feedback_2011, altug_feedback_2014, wagner_new_2020, fong_asymptotic_2015, fong_tight_2017} and multiple other applications. For identification via channels, Watanabe \cite{watanabe_minimax_2022} derived the second-order coding rate for the DMC. This result was extended to the Gaussian channel by Liu et al. \cite{liu_second-order_2024-1}. For other identification scenarios, the second-order rates are still an open problem, often due to the lack of a strong converse \cite{bruche_converse_2025-1}. 

In this work, we investigate the second-order coding rate of identification with noiseless feedback. For deterministic encoding, we prove an achievable second-order rate, which we show to be optimal for the class of \emph{simple-dispersion channels}, which includes the most common channel models like the binary symmetric channel (BSC) and binary erasure channel (BEC). For randomized encoding, we prove the optimal second-order coding rate for all DMCs. To prove the achievability, we employ a two-phase code construction as introduced in \cite{ahlswede_identification_1989-1}. However, the original achievability proof relies heavily on typical sets and is therefore not suitable for a second-order analysis. We introduce a new receiver operation to overcome this issue, and we shorten the length of the second phase to $O(\log n)$ (which was already shown in \cite{verdu_explicit_1993} to be capacity-achieving) and carefully analyze the resulting error probabilities in order to achieve the best possible second-order rates. For the converse, the feedback introduces a possible dependence between the input symbols, which is why we use martingale-based techniques, inspired by the work on transmission with feedback \cite{wagner_new_2020} and adapt them to the case of identification with feedback. For both deterministic and randomized identification, we also provide interesting insights into the second-order behavior of channels with certain symmetry properties.
We note that both our results for deterministic and for randomized identification with feedback cannot be generalized to the Gaussian channel since the identification-feedback capacity of the Gaussian channel is infinite \cite{labidi_identification_2021}, making a meaningful definition of a second-order rate impossible.

The remainder of the paper is organized as follows: Section \ref{sec:det} contains the model and our result for deterministic identification, while Section \ref{sec:rand} contains those for randomized identification. In Section \ref{sec:num}, we illustrate our results for some example channels. Sections \ref{sec:proof-det} and \ref{sec:proof-rand} contain our proofs. 

\subsection{Notation}
For an arbitrary set $\mathcal{X}$, $\mathcal{P}(\mathcal{X})$ denotes the set of all probability distributions on $\mathcal{X}$. For $P \in \mathcal{P}(\mathcal{X})$, the entropy is denoted by $H(P)$. The marginal output distribution induced by an input distribution $P$ on a channel $W$ is denoted by
$PW(y) = \sum_{x\in \mathcal{X}}P(x)W(y|x).$ The joint distribution of input and output is denoted by $(P \times W)(x, y) = P(x)W(y|x)$.
We use $Q(\cdot)$ to denote the Gaussian tail distribution function {and we use the standard Landau symbols $o(\cdot)$ and $O(\cdot)$}. We use the notation $P_{x^n}(x)$ to denote the type of a sequence $x^n$. All $\log$ and $\exp$ functions are base 2.
\section{Deterministic Identification with Feedback}\label{sec:det}
\subsection{System Model and Preliminaries}
We consider a discrete memoryless channel (DMC) $W$ with input alphabet $\mathcal{X}$ and  output alphabet $\mathcal{Y}$. We assume that all output symbols are reachable, i.e., for each $y \in \mathcal{Y}$ there exists an $x \in \mathcal{X}$ such that $W(y|x) > 0$ (If a channel does not fulfill that assumption, we can replace it by an equivalent channel where all unreachable output symbols are removed from $\mathcal{Y}$.) A noiseless feedback link without rate constraint from the receiver to the transmitter is available, which makes available the received symbol $Y_{t-1}$ to the transmitter at time $t$. The set of deterministic feedback encoding functions is defined by
\begin{align}
    \mathcal{F}_n := \{\vec{f} = [f^1, f^2, ..., f^n]\, :\,&  f^1 \in \mathcal{X}, \nonumber\\
    &f^t: \mathcal{Y}^{t-1} \to \mathcal{X}, t = 2, ..., n \}.
\end{align}
The receiver is interested in a specific message $i' \in [N_n]$, which is unknown to the transmitter. The transmitter encodes his message $i$ in such a way that the receiver will be able to reliably identify whether $i = i'$ or $i \neq i'$ (see Fig. \ref{fig:identification-feedback}). 
\begin{figure}
    \centering
    \tikzstyle{block} = [draw, fill=SkyBlue!20, rectangle,
minimum height=2em, minimum width=2.5em]

\tikzstyle{input} = [coordinate]
\tikzstyle{sum} = [draw, circle,inner sep=0pt, minimum size=2mm,  thick]

\scalebox{.75}{
\tikzstyle{arrow}=[draw,->]
\begin{tikzpicture}[auto, node distance=2cm,>=latex']
\node[] (M) {${i}\in \{1, ... N\}$};
\node[block,right=.5cm of M] (enc) {Encoder};
\node[block, right=2.5cm of enc] (channel) {DMC};
\node[block, right=1cm of channel] (dec) {Decoder};
\node[align=left,right=.5cm of dec] (iprime) {Was ${i^\prime}$\\ sent?};
\node[input,right=.5cm of channel] (t1) {};
\node[input,below=1cm of t1] (t2) {};
\draw[->,thick] (M) -- (enc);
\draw[->,thick] (enc) --node[above]{ $X_t={f}^t_{i}(Y^{t-1})$} (channel);
\draw[->,thick] (channel) --node[above]{$Y_t$} (dec);
\draw[->,thick] (dec) -- (iprime);
\draw[->,thick] ([yshift=6mm]dec.north) node[above] {$i'$} -- (dec.north);
\draw[-,thick] (t1) -- (t2);
\draw[->,thick] (t2) -| (enc);
\end{tikzpicture}}{
    \caption{System model of deterministic identification with noiseless feedback}
    \label{fig:identification-feedback}}
\end{figure}
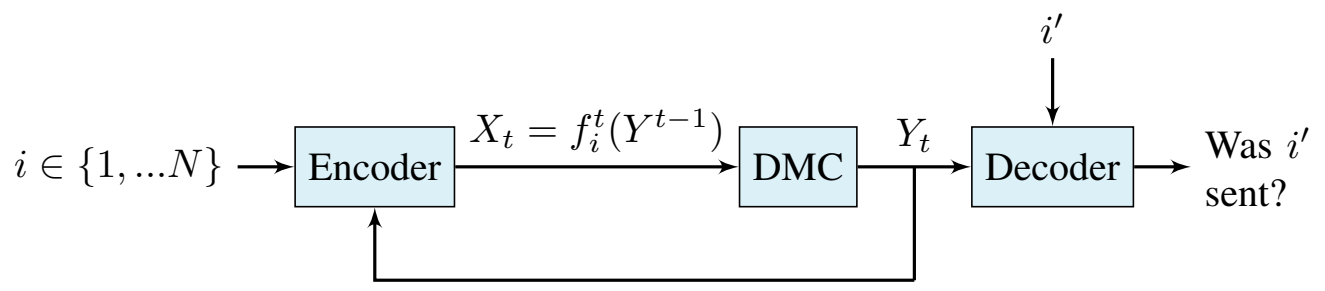
\begin{definition}\label{def:idf-det-code}
    A \emph{deterministic $(n, N_n, \lambda_1, \lambda_2)$-identification-feedback (IDf) code} for a DMC $W$ is a family $\left\{(\vec{f}_i, \mathcal{D}_i): i \in [N_n]\right\}$
    consisting of encoding functions $\vec{f}_i \in \mathcal{F}_n$ and decoding sets $\mathcal{D}_i \subset \mathcal{Y}^n$,
    such that the errors of the first kind and second kind satisfy
    \begin{align}
        \lambda_{1}^{(i)} := W^n(\mathcal{D}^c_{i}| \vec{f}_i) \leq \lambda_1, \quad
         \lambda_{2}^{(i, j)} := W^n(\mathcal{D}_{j}|\vec{f}_i) \leq \lambda_2
    \end{align}
    for all $i,j \in [N_n]$ with $i \neq j$.
\end{definition}
\begin{definition}\label{def:idf-det-logN}
    For $0 < \lambda_1, \lambda_2 < 1$, $\lambda_1 + \lambda_2 < 1$, we define the \emph{optimal code size} of deterministic identification via channel $W$ with feedback by
    \begin{align}
        N_{\mathsf{IDf}}^*(n, \lambda_1, \lambda_2|W) := \max\{N:& \,\exists\, (n, N, \lambda_1, \lambda_2)\text{-IDf code} \nonumber \\
        &\text{for channel }W\}.
    \end{align}
\end{definition}
\begin{definition}\label{def:idf-det-capacity}
    For given $0 < \lambda_1, \lambda_2 < 1$, $\lambda_1 + \lambda_2 < 1$, the \emph{deterministic $(\lambda_1, \lambda_2)$-identification-feedback (IDf) capacity} of a DMC $W$ is defined as
    \begin{align}
        C_{\mathsf{IDf}}(\lambda_1, \lambda_2|W) := \liminf_{n \to \infty} \frac{\log \log N_{\mathsf{IDf}}^*(n, \lambda_1, \lambda_2|W)}{n}.
    \end{align}
\end{definition}
The definition of the rate as double logarithmic is motivated by the observation that $N_{\mathsf{IDf}}^*(n, \lambda_1, \lambda_2|W)$ grows double exponentially. The IDf capacity was characterized by Ahlswede and Dueck. 
\begin{theorem}[{Ahlswede and Dueck \cite{ahlswede_identification_1989-1}}]\label{thm:idf-det-capacity}
For $\lambda_1, \lambda_2 < \frac{1}{2}$, the deterministic $(\lambda_1, \lambda_2)$-identification-feedback capacity of a DMC $W$ with transmission capacity $C(W)>0$ is given by
    \begin{align}
        C_{\mathsf{IDf}}(\lambda_1, \lambda_2|W) = \max_{x \in \mathcal{X}} H(W(\cdot|x)).
    \end{align}
\end{theorem}
As we will see later, this result can be generalized to the case $\lambda_1 + \lambda_2 < 1$. In contrast to the transmission or ID capacity without feedback, the IDf capacity is not characterized by the mutual information, but by the conditional channel output entropy, meaning that more channel noise increases the capacity as long as the transmission capacity is still positive. This is due to the availability of noiseless feedback and the fact that identification capacities are essentially limited by how much common randomness can be generated between transmitter and receiver. In the case of Gaussian channels, this makes the IDf capacity infinitely large \cite{labidi_identification_2021}. 

The second-order behaviour is characterized by the second-order coding rate, which we define analogously to the case of identification without feedback \cite{watanabe_minimax_2022}.
\begin{definition}\label{def:idf-det-second-order}
    For given $0 < \lambda_1, \lambda_2 < 1$, $\lambda_1 + \lambda_2 < 1$, the \emph{deterministic second-order $(\lambda_1, \lambda_2)$-identification-feedback {(IDf)} capacity} of a {DMC} $W$ is defined as
    \begin{align}
        &L_{\mathsf{IDf}}(\lambda_1, \lambda_2|W) \nonumber\\
        &:= \liminf_{n \to \infty} \frac{\log \log N_{\mathsf{IDf}}^*(n, \lambda_1, \lambda_2|W) - nC_{\mathsf{IDf}}(\lambda_1, \lambda_2|W)}{\sqrt{n}}
    \end{align}
    and the \emph{deterministic second-order $\lambda_1$-{IDf} capacity} of a {DMC} $W$ is defined as
    \begin{align}
        L_{\mathsf{IDf}}(\lambda_1|W) := \lim_{\lambda_2 \to 0} L_{\mathsf{IDf}}(\lambda_1, \lambda_2|W).
    \end{align}
\end{definition}
Define the set of IDf-capacity-achieving input symbols as
\begin{align}
    \mathcal{X}^*_W := \{x \in \mathcal{X}: H(W(\cdot|x)) \geq H(W(\cdot|\tilde{x})) \;\forall \tilde{x} \in \mathcal{X}\} \label{eq:optimal-input-set-det}
\end{align}  
and the minimum and maximum IDf-dispersions as
\begin{align}
    V_{\mathsf{IDf}}^{\min}(W) &= \min_{x \in \mathcal{X}^*_W}\Var\left[\log W(Y|x)\right], \\
    V_{\mathsf{IDf}}^{\max}(W) &= \max_{x \in \mathcal{X}^*_W}\Var\left[\log W(Y|x)\right].
\end{align}

\begin{remark}
    In this work, we use both the notion of second-order coding rates as defined in Def. \ref{def:idf-det-second-order} and of channel dispersions. The notion of second-order coding rates was first introduced by Hayashi \cite{hayashi_information_2009} and captures the whole second-order behavior. The channel dispersion was first defined by Polyanskiy, Poor and Verdú \cite{polyanskiy_channel_2010} and captures only the channel-dependent part of the second-order rate, separate from the part that depends on the error probability. We will use the term \textit{dispersion} purely in a non-operational way; our results are expressed in terms of second-order coding rates.
\end{remark}

\begin{definition}\label{def:idf-det-simple-dispersion}
    A DMC $W$ with $V_{\mathsf{IDf}}^{\min}(W) > 0$ is called \emph{IDf-simple-dispersion} if $V_{\mathsf{IDf}}^{\min}(W) = V_{\mathsf{IDf}}^{\max}(W) =: V_{\mathsf{IDf}}(W)$. Otherwise, it is called \emph{IDf-compound-dispersion}.
\end{definition}
Note that the class of simple-dispersion channels includes all weakly symmetric channels, which will also be considered in this work. 
\begin{definition}[{\cite[p. 190]{cover_elements_2006}}]
    A DMC $W$ is called \textit{weakly symmetric} if the probability vectors $W(\cdot|x)$, $x \in \mathcal{X}$, are permutations of each other and the sums $\sum_x W(y|x)$ are independent of $y$.
\end{definition}

\begin{example}\label{ex:compound-dispersion}
    Consider a DMC with input alphabet $\mathcal{X} = \{0, 1\}$ and output alphabet $\mathcal{Y} = \{0, 1, 2\}$. The channel is given by the matrix
    \begin{align}
        \begin{pmatrix}
            p & \frac{1-p}{2} & \frac{1-p}{2} \\
            q & 1-q & 0
        \end{pmatrix}.
    \end{align}
    We choose $p$ and $q$ such that 
    \begin{align}
        H\left(p, \frac{1-p}{2}, \frac{1-p}{2}\right) = H(q, 1-q), \label{eq:idf-compound-dispersion-channel-entropy}
    \end{align}
    i.e., both inputs result in the same output entropy and are thus both IDf-capacity achieving. A numerical search yields that $p = 0.8$ and $q \approx 0.337$ is a solution to \eqref{eq:idf-compound-dispersion-channel-entropy}. Thus, for this choice of $p$ and $q$, both inputs $x = 0$ and $x = 1$ are capacity-achieving. However, the corresponding dispersions differ: We have
    $\Var\left[\log{W(Y|x = 0)}\right] \approx 1.44$ and $\Var\left[\log {W(Y|x = 1)}\right] \approx 0.213.$
    Therefore, the channel is IDf-compound-dispersion.
\end{example}

\subsection{Main Result}
\begin{theorem}\label{thm:second-order-idf} Let $W$ be a DMC with transmission capacity $C(W)>0$ and let
\begin{align}
    V_{\mathsf{IDf}}(\lambda_1, W) := \begin{cases}
        V_{\mathsf{IDf}}^{\min}(W), & 0 < \lambda_1 \leq \frac{1}{2}, \\
        V_{\mathsf{IDf}}^{\max}(W), & \frac{1}{2} < \lambda_1 < 1.
    \end{cases}
\end{align}
For $\lambda_1+\lambda_2 < 1$, the deterministic $(\lambda_1, \lambda_2)$-identification-feedback capacity of $W$ is given by
    \begin{align}
        C_{\mathsf{IDf}}(\lambda_1, \lambda_2|W) = \max_{x \in \mathcal{X}} H(W(\cdot|x)) \label{eq:idf-det-capacity}
    \end{align}
    and the deterministic second-order $\lambda_1$-identification-feedback capacity of $W$ is bounded by
\begin{align}
    L_{\mathsf{IDf}}(\lambda_1|W) \geq  
            - \sqrt{V_{\mathsf{IDf}}(\lambda_1, W)}Q^{-1}(\lambda_1). \label{eq:idf-det-second-order-rate}
\end{align}
Equality holds if the channel is IDf-simple dispersion.
\end{theorem}

The proof details can be found in Section \ref{sec:proof-det}. We adapt the achievability proof from \cite{ahlswede_identification_1989-1} to the second-order regime.
For the converse proof, we extend the fundamental argument behind the proof in \cite{ahlswede_identification_1989} to the case $\lambda_1 + \lambda_2 < 1$ and conduct an analysis based on martingale techniques, inspired by the work on transmission with feedback \cite{wagner_new_2020}.

\begin{corollary} \label{cor:idf-det-symmetric} For all weakly symmetric channels, we have $\mathcal{X}_W^* = \mathcal{X}$ and $V_{\mathsf{IDf}}(W) = V(W)$
    where $V(W)$ denotes the transmission dispersion of the DMC $W$.
\end{corollary}
\begin{IEEEproof}
The weak input symmetry directly implies that for all $(x,x') \in \mathcal{X} \times \mathcal{X}$,
$H(W(\cdot|x)) = H(W(\cdot|x'))$ and $\Var\left[\log W(Y|x)\right] = \Var\left[\log W(Y|x')\right].$
This implies that all input symbols achieve the capacity, which proves that $\mathcal{X}_W^* = \mathcal{X}$.

Furthermore, it is known that for weakly symmetric channels, the transmission capacity is achieved by a uniform distribution, and the capacity-achieving output distribution is also uniform \cite[p. 190]{cover_elements_2006}. Thus,
    $\mathbb{E}\left[\log PW(Y)\right] = \log\frac{1}{|\mathcal{Y}|},$
    $\Var\left[\log PW(Y)\right] = 0$ and
    $\text{Cov}\left[\log W(Y|X), \log PW(Y)\right] =0.$
Thus, we conclude
    \begin{IEEEeqnarray}{rCl}
        V(W) &=& \Var\left[\log \frac{W(Y|X)}{PW(Y)}\right] \nonumber\\
        &=& \sum_{x \in \mathcal{X}} P(x) \Var\left[\log W(Y|x)\right] \\
        &=& \Var\left[\log W(Y|x')\right] = V_{\mathsf{IDf}}(W),
    \end{IEEEeqnarray}
    with an arbitrary $x' \in \mathcal{X}$, where the last two steps follow from the weak symmetry.
\end{IEEEproof}
\section{Randomized Identification with Feedback}\label{sec:rand}
\subsection{System Model and Preliminaries}
By randomized identification, we mean that the encoder has its own unlimited source of local randomness which can be used in the encoding process. We differentiate the terms by using the abbreviations \textit{IDf} for deterministic identification and \textit{IDF} for randomized identification.

\begin{definition}\label{def:idf-rand-code}
    A \emph{randomized $(n, N_n, \lambda_1, \lambda_2)$-identification-feedback (IDF) code} for a DMC $W$ is a family $\left\{(Q_F(\cdot|i), \mathcal{D}_i): i \in [N_n]\right\}$
    consisting of stochastic encoding functions $Q_F(\cdot|i) \in \mathcal{P}(\mathcal{F}_n)$ and decoding sets $\mathcal{D}_i \subset \mathcal{Y}^n$,
    such that the errors of the first and second kind satisfy
    \begin{align}
        \lambda_{1}^{(i)} &:= \sum_{\vec{g}\in \mathcal{F}_n} Q_F(\vec{g}|i)W^n(\mathcal{D}^c_{i}| \vec{g}) \leq \lambda_1, \\
         \lambda_{2}^{(i, j)} &:= \sum_{\vec{g}\in \mathcal{F}_n} Q_F(\vec{g}|i) W^n(\mathcal{D}_{j}|\vec{g}) \leq \lambda_2
    \end{align}
    for all $i,j \in [N_n]$ with $i \neq j$.
\end{definition}
We define $N_{\mathsf{IDF}}^*(n, \lambda_1, \lambda_2|W)$, $C_{\mathsf{IDF}}(\lambda_1, \lambda_2|W)$, $L_{\mathsf{IDF}}(\lambda_1, \lambda_2|W)$, and $L_{\mathsf{IDF}}^*(\lambda_1|W)$ analogously to the deterministic case. 
Again, the IDF capacity was {established} by Ahlswede and Dueck.
\begin{theorem}[{Ahlswede and Dueck \cite{ahlswede_identification_1989-1}}]\label{thm:idf-rand-capacity} For $\lambda_1, \lambda_2 < \frac{1}{2}$, the randomized $(\lambda_1, \lambda_2)$-identification-feedback capacity of a DMC $W$ with transmission capacity $C(W)>0$ is given by
    \begin{align}
        C_{\mathsf{IDF}}(\lambda_1, \lambda_2|W) = \max_{P \in \mathcal{P}(\mathcal{X})} H(P W).
    \end{align}
\end{theorem}
Let the set of IDF-capacity-achieving input distributions be given by
\begin{align}
    \Phi_W^* := \{P \in \mathcal{P}(\mathcal{X}): H(PW) \geq H(\tilde{P} W) \;\forall \tilde{P} \in \mathcal{P}(\mathcal{X})\}. \label{eq:optimal-input-set-rand}
\end{align}  
Then we the minimum and maximum IDF-dispersions are
\begin{align}
    V_{\mathsf{IDF}}^{\min}(W) &= \min_{P \in  \Phi_W^*}\Var\left[\log {P W(Y)}\right], \\
    V_{\mathsf{IDF}}^{\max}(W) &= \max_{P \in  \Phi_W^*}\Var\left[\log {P W(Y)}\right].
\end{align}

\subsection{Main Result}
For randomized identification with feedback, we can fully characterize the IDF-capacity since all channels are IDF-simple dispersion, as the following Theorem shows.

\begin{theorem}\label{thm:idf-rand-simple-dispersion}
For any DMC $W$, the following equality always holds:
    \begin{align}
        V_{\mathsf{IDF}}^{\min}(W) &= V_{\mathsf{IDF}}^{\max}(W) =: V_{\mathsf{IDF}}(W)
    \end{align}
\end{theorem}
The proof can be found in Section \ref{sec:proof-unique-dispersion}.
With this result, we can obtain the following full characterization of the second-order coding rate of randomized identification with feedback:
\begin{theorem}\label{thm:second-order-idf-rand} Let $W$ be a DMC with transmission capacity $C(W)>0$. For $\lambda_1+\lambda_2 < 1$, the randomized $(\lambda_1, \lambda_2)$-identification-feedback capacity of $W$ is given by
    \begin{align}
        C_{\mathsf{IDF}}(\lambda_1, \lambda_2|W) = \max_{P \in \mathcal{P}(\mathcal{X})} H(PW)
    \end{align}
    and the randomized second-order $\lambda_1$-identification-feedback capacity of $W$  is given by
\begin{align}
    L_{\mathsf{IDF}}(\lambda_1|W) =  
            - \sqrt{V_{\mathsf{IDF}}(W)}Q^{-1}(\lambda_1)
\end{align}
\end{theorem}
The proof can be found in Section \ref{sec:proof-rand}. 

Again, we can find a direct consequence for channels with certain symmetry properties. In the case of randomized identification with feedback, we find the remarkable result that the second-order coding rate is zero for all channels with a uniform capacity-achieving output distribution, which includes all weakly symmetric channels.
\begin{corollary} \label{cor:idf-rand-symmetric}For all DMCs $W$, we have $V_{\mathsf{IDF}}(W) = 0$ iff
the IDF-capacity-achieving output distribution satisfies 
\begin{align}
    P^*W(y) = \frac{1}{|\mathcal{Y}|} \text{ for all } y \in \mathcal{Y}.
\end{align}
\end{corollary}

\begin{IEEEproof} The IDF-dispersion of a channel is defined as
    \begin{align}
        &V_{\mathsf{IDF}}(W) = \Var\left[\log {P^*W(Y)}\right] \\
        &= \sum_{y\in \mathcal{Y}} P^*W(y)  \left(\log {P^*W(y)} - \expect{\log {P^*W(Y)}}\right)^2. \label{eq:idf-rand-dispersion-zero}
    \end{align}
    Since $P^*W(y)$ must be positive for some $y \in \mathcal{Y}$, \eqref{eq:idf-rand-dispersion-zero} is zero iff $\log P^*W(y) = \expect{\log {P^*W(Y)}}$. This requires all nonzero $P^*W(y)$ to be equal, which requires $P^*W(y) = \frac{1}{|\mathcal{Y}|}$.
\end{IEEEproof}

Note that a uniform output distribution $P^*W$ also means that the IDF-capacity is $C_{\mathsf{IDF}} = H(P^*W) = \log |\mathcal{Y}|$, which is independent of the channel properties. Thus, Corollary \ref{cor:idf-rand-symmetric} tells us that this remarkably large capacity is also approached very quickly, with only higher-order terms of $O(\log n)$ remaining.

\section{Numerical Results}\label{sec:num}
\begin{figure}
    \centering
    \beginpgfgraphicnamed{example1_second_order_rates_plot}
\begin{tikzpicture}
	\begin{axis}[
        width=\linewidth,
        height=7cm,
		xlabel=$n$,
		ylabel=$C+L/\sqrt{n}$,
		xmin=100,
		xmax=2000,
		ymin=0.75,
		ymax=1.23,
		xmajorgrids=true,
		ymajorgrids=true,
		legend columns=2,
		legend style={at={(0.5,-0.2)},
		              anchor=north,
		              /tikz/column 2/.style={column sep=5pt,},
		              draw=none},
		legend cell align={left},
		style={font=\footnotesize}]

		\addplot[color=plot0, style={very thick, dashed}]
		table[x=n,y=idf_nonoptimal,comment chars={\#}]
		{./figures/example1_second_order_rates.dat};
		\addlegendentry{IDf, high-dispersion input}

		\addplot[color=plot0, style={very thick}]
		table[x=n,y=idf_optimal,comment chars={\#}]
		{./figures/example1_second_order_rates.dat};
		\addlegendentry{IDf, small-dispersion input}

		\addplot[color=plot2, style={very thick}]
		table[x=n,y=idf_timid_bold,comment chars={\#}]
		{./figures/example1_second_order_rates.dat};
		\addlegendentry{IDf, timid/bold}

		\addplot[color=plot1, style={very thick, dash pattern=on 6pt off 2pt on 1pt off 2pt}]
		table[x=n,y=idf,comment chars={\#}]
		{./figures/example1_second_order_rates.dat};
		\addlegendentry{IDF}

		\addplot[color=plot0, style={very thick, dotted}]
		table[x=n,y=idf_capacity,comment chars={\#}]
		{./figures/example1_second_order_rates.dat};
		\addlegendentry{IDf capacity}

		\addplot[color=plot1, style={very thick, dotted}]
		table[x=n,y=idf_randomized_capacity,comment chars={\#}]
		{./figures/example1_second_order_rates.dat};
		\addlegendentry{IDF capacity}

	\end{axis}
\end{tikzpicture}
\endpgfgraphicnamed
    \caption{Gaussian approximations of the IDf- and IDF-rates for the compound-dispersion channel from example \ref{ex:compound-dispersion} with $\lambda_1 = 0.1$. }
    \label{fig:second-order-rate-plot}
\end{figure}
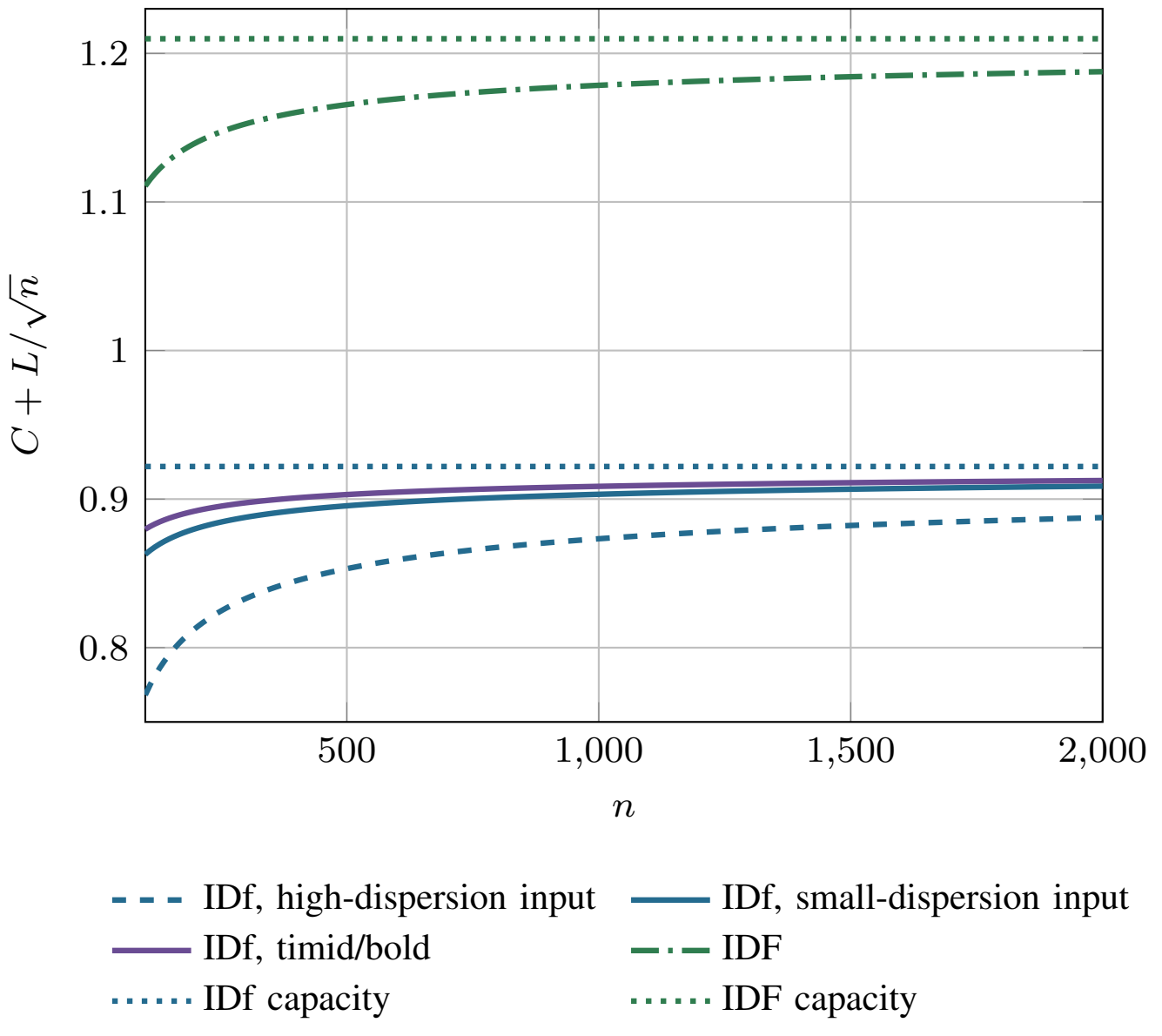

To illustrate our results, we consider the channel from Example \ref{ex:compound-dispersion}, which is compound-dispersion, i.e., it does not fulfill the simple-dispersion assumption. This channel has the capacities 
$C_{\mathsf{IDf}}(\lambda_1, \lambda_2|W) \approx 0.92$ bits/channel use and $C_{\mathsf{IDF}}(\lambda_1, \lambda_2|W) \approx 1.21$ bits/channel use.
In Fig. \ref{fig:second-order-rate-plot}, we show the normal approximation of the rate 
$R \approx C + \frac{L}{\sqrt{n}}$,
where $C$ stands for the corresponding capacity expression and $L$ stands for the corresponding second-order capacity. To illustrate the effect of the non-simple dispersion behavior, we plot the achievable rates for IDf using both capacity-achieving input distributions and observe that the dispersion-minimizing distribution is noticeably better at small blocklengths. For larger blocklengths, this difference gets smaller, which is expected since both rates eventually converge towards the same capacity. However, even the curve with the second-order optimal input distribution can not be considered an upper bound on the achievable rates since our Theorem \ref{thm:second-order-idf} does not have a converse for compound-dispersion channels. 
In fact, Wagner et al. \cite{wagner_new_2020} have shown for the case of message transmission that a strategy called \textit{timid/bold coding} can further improve the rates. This strategy can also be applied for deterministic identification. In short, the encoder can adaptively switch between the two capacity-achieving input symbols $x= 0$ and $x=1$: if the instantaneous amount of randomness at the receiver is smaller than a certain threshold, the transmitter is "bold" and uses the high-variance input symbol $x=0$. Otherwise, it is "timid" and uses $x=1$. The resulting second-order rate is
\begin{equation*}
\begin{split}
L_{\mathrm{IDf}}(\lambda_1|W)
\ge
\begin{cases}
-\sqrt{V_{\mathrm{IDf}}^{\min}(W)}
\,Q^{-1}\!\left(
\frac{\lambda_1(1+\beta_{\mathrm{IDf}})}
     {2\beta_{\mathrm{IDf}}}
\right),
\\
\hfill
0<\lambda_1
\le
\frac{\beta_{\mathrm{IDf}}}
     {1+\beta_{\mathrm{IDf}}},
\\[2ex]
-\sqrt{V_{\mathrm{IDf}}^{\max}(W)}
\,Q^{-1}\!\left(
\frac{
\lambda_1(1+\beta_{\mathrm{IDf}})
+1-\beta_{\mathrm{IDf}}
}{2}
\right),
\\
\hfill
\frac{\beta_{\mathrm{IDf}}}
     {1+\beta_{\mathrm{IDf}}}
<\lambda_1<1,
\end{cases}
\end{split}
\end{equation*}
where 
    $\beta_{\mathrm{IDf}}:= \sqrt{\frac{V_{\mathrm{IDf}}^{\min}(W)}          {V_{\mathrm{IDf}}^{\max}(W)}}$.
This result can be deduced using the same techniques and steps as in \cite[Section IV]{wagner_new_2020}, which is why we refrain from giving a proof here. In Fig. \ref{fig:second-order-rate-plot}, it can be observed that the rates that can be achieved by timid/bold coding are larger than by just using the optimal input symbol all the time. However, there is no converse which shows that timid/bold coding is optimal, leaving the question whether there is an even better achievability strategy open.

For randomized IDF, we know the optimal input distribution. Therefore, only one curve is shown in Fig. \ref{fig:second-order-rate-plot}, and timid/bold coding is not possible. Since we have a matching converse, this curve shows the second-order approximation of the optimal rate.

In a second example, we consider the binary erasure channel with erasure probability $\epsilon = 0.25$, which has the capacities 
\begin{align}
    C_{\mathsf{IDf}}(\lambda_1, \lambda_2|W) &= \hb(\epsilon) \approx 0.81, \text{ and} \\
    C_{\mathsf{IDF}}(\lambda_1, \lambda_2|W) &= \hb(\epsilon) + 1 - \epsilon \approx 1.56.
\end{align}
This channel is simple-dispersion, but not weakly symmetric and the IDF-capacity-achieving output distribution is not uniform: the probability of the erasure output symbol is always $W(\mathsf{e}|x) = \epsilon$ for both $x = 0,1$. Thus, the assumptions for both Corollaries \ref{cor:idf-det-symmetric} and \ref{cor:idf-rand-symmetric} are not fulfilled. Indeed, we have
\begin{align}
    V_{\mathsf{IDf}}(W) &= \epsilon(1-\epsilon) \left(\log \frac{1-\epsilon}{\epsilon}\right)^2 \approx 0.47, \text{ and} \\
    V_{\mathsf{IDF}}(W) &= \epsilon(1-\epsilon) \left(\log \frac{1-\epsilon}{2\epsilon}\right)^2 \approx 0.06.
\end{align}
Especially for the randomized case, the very small dispersion is noticeable since the rates are already very close to capacity at small blocklengths. Note that since the channel is simple-dispersion, timid/bold coding cannot be applied and the second-order rates for both IDf and IDF are already optimal.

\section{Proof of Theorem \ref{thm:second-order-idf}}\label{sec:proof-det}
\subsection{Achievability (Nonzero variance)}\label{sec:idf-det-achievability}
We use the two-phase construction from \cite{ahlswede_identification_1989-1} and adapt the construction to make it suitable for a second-order analysis. Assume $V_{\mathsf{IDf}}(\lambda_1, W) > 0$. Let $\gamma > 0$ be sufficiently large, $\nb = \lceil\gamma \log \ntot\rceil$ and $\na = \ntot - \nb$. Consider a maximal-error $(\nb, M'', \mu_{\nb}'')_{\text{max}}$-transmission code
    \begin{align}
        \mathcal{C}'' = \{(\vec{u}_k'', \mathcal{D}_k''): \, k \in  [M'']\}
    \end{align}
    with rate $R_c'' = \frac{\log M''}{\nb}$.
 Choose a set of hash functions
    \begin{align}
        F_i: \mathcal{D}^* \subset\mathcal{Y}^{\na} \to [M''], \; i = 1, ..., N,
    \end{align}
    where we refer to the output as \textit{tag}. Choose an $x^* \in \mathcal{X}^*_W$ that achieves $V_{\mathsf{IDf}}(\lambda_1, W)$. The set $\mathcal{D}^*$ is defined as
    \begin{align}
        \mathcal{D}^* &:= \left\{y^{\na} \in \mathcal{Y}^{\na}: \, \log \frac{1}{W^{\na}(y^{\na}|x^{*{\na}})} \geq \log \rho_{\na}\right\}, \label{eq:idf-ach2-set-d}
    \end{align}
    where \begin{align}
        \log \rho_{\na} &= {\na}H(W(\cdot | x^{*})) - \sqrt{{\na}V_{\mathsf{IDf}}(\lambda_1, W)}Q^{-1}(\mu_{\na}'), \label{eq:idf-ach-proof-log-rho}
    \end{align}
    and $\mu_{\na}'$ is to be specified later.
To transmit message $i \in [N]$, 
during the first ${\na}$ transmissions, the transmitter sends the symbol $x^*$.
Then it calculates the tag $j$ using the $i$-th hash function and the received feedback $y^{\na}$: $j = F_i(y^{\na})$. If $y^{\na} \notin \mathcal{D}^*$, it chooses $j = 1$ instead.
It then transmits $\vec{u}''_{j}$ over the channel during the last ${\nb}$ transmission times.
The receiver is interested in message $i' \in \{1, ..., N\}$. 
After receiving the first ${\na}$ symbols $y^{\na}$, the receiver checks whether $y^{\na} \in \mathcal{D}^*$. If not, the receiver rejects $i'$. 
Otherwise, it calculates the tag $F_{i'}(y^{\na})$.
Based on the last $\nb$ received symbols, it decodes an estimate of the received tag $\hat{j}$.
If $F_{i'}(y^{\na}) = \hat{j}$, it declares that $i'$ was sent.
Thus, the decoding sets are
\begin{align}
    \mathcal{D}(F_i) = \bigcup_{y^{\na} \in \mathcal{D}^*} \{y^{\na}\} \times \mathcal{D}''_{F_i(y^{\na})}, \; \forall i \in [N].
\end{align}
For our construction of an identification-feedback code, we choose the family of hash functions $\{F_i: i \in [N]\}$ independently and uniformly at random, such that we have random hash functions $\bar{F}(i)(y^{\na}) \sim \text{Unif}( [M''])$.

\paragraph{Error of the first kind.} For every realization $F_i$ of $\bar{F}_i$, we have
    \begin{align}
        \lambda_{1}^{(i)} &= W^{\ntot}(\mathcal{D}(F_i)^c|{f}_i) \nonumber\\
        &\leq  W^{\na}((\mathcal{D}^{*})^c|x^{*\na}) + W^{{\nb}}(\mathcal{D}_{F_i(y^{\na})}''^c|\vec{u}_{F_i(y^{\na})}'') \nonumber\\ 
        &\leq \Pr\left[ \log \frac{1}{W^{\na}(Y^{\na}|x^{*{\na}})} < \log \rho_{\na}\right] + \mu_{\nb}'', \label{eq:idf-ach2-first-kind-error-bound-code2} \\
        &\leq \mu_{\na}' + \delta_{\na} + \mu_{\nb}''{,} \label{eq:idf-ach2-first-kind-error}
    \end{align}
where \eqref{eq:idf-ach2-first-kind-error-bound-code2} follows from the properties of the code $\mathcal{C}''$ and \eqref{eq:idf-ach2-first-kind-error} follows from the Berry-Esseen Theorem \cite[Theorem 44]{polyanskiy_channel_2010}. We abbreviate the Berry-Esseen remainder as $\delta_{\na}$.

 \paragraph{Error of the second kind.} We fix a realization $F_{i_1}$ of $\bar{F}_{i_1}$ and let $\bar{F}_{i_2}$ be random. Denote the error of erroneously identifying $i_2$ when $i_1$ is sent with $\bar{\lambda}^{(i_1, i_2)}_2$ when $\bar{F}_{i_2}$ is random. We have
    \begin{align}
        \bar{\lambda}^{(i_1, i_2)}_2 =& W^{\ntot}(\mathcal{D}(\bar{F}_{i_2})|{f}_{i_1})\nonumber \\
        \leq& \sum_{y^{\na} \in \mathcal{D}^*}W^{\ntot}\left(\{y^{\na}\} \times \mathcal{D}''_{\bar{F}_{i_2}(y^{\na})}\mid {f}_{i_1}\right) \nonumber\\
        \leq& \!\!\!\!\!\!\!\!\!\!\sum_{\substack{y^{\na} \in \mathcal{D}^*\\F_{i_1}(y^{\na}) = \bar{F}_{i_2}(y^{\na})}}\!\!\!\!\!\!\!\!\!\!\!\!\!\!W^{\na}(y^{\na}|x^{*{\na}}) +\!\!\!\!\!\!\!\!\!\!\!\!\!\!\sum_{\substack{y^{\na} \in \mathcal{D}^*\\F_{i_1}(y^{\na}) \neq \bar{F}_{i_2}(y^{\na})}} \!\!\!\!\!\!\!\!\!\!\!\!\!\!W^{\na}(y^{\na}|x^{*{\na}})\mu_{\nb}''\nonumber\\
        \leq& \!\!\!\!\!\!\!\!\!\!\sum_{\substack{y^{\na} \in \mathcal{D}^*\\F_{i_1}(y^{\na}) = \bar{F}_{i_2}(y^{\na})}}\!\!\!\!\!\!\!\!\!\!W^{\na}(y^{\na}|x^{*{\na}}) + \mu_{\nb}''. \label{eq:idf-ach-intersectionset}
    \end{align}
    Now, we analyze the first term in \eqref{eq:idf-ach-intersectionset}. Define 
    \begin{align}
        \psi_{y^{\na}} = \psi_{y^{\na}}(\bar{F}_{i_2}) = \begin{cases}
            1 &\text{if } \bar{F}_{i_2}(y^{\na}) = F_{i_1}(y^{\na}), \\
            0 &\text{otherwise}.
        \end{cases}
    \end{align}
    The random variables $\{\psi_{y^{\na}}\: y^{\na}\in \mathcal{D}^*\}$ are independent and we have for all $y^{\na} \in \mathcal{D}^*$
    \begin{align}
        \mathbb{E}[\psi_{y^{\na}}] &= \Pr[\bar{F}_{i_2}(y^{\na}) = F_{i_1}(y^{\na})] = \frac{1}{M''}.
    \end{align}
    With this definition, we can write
    \begin{align}
        \sum_{\substack{y^{\na} \in \mathcal{D}^*\\F_{i_1}(y^{\na}) = \bar{F}_{i_2}(y^{\na})}}\!\!\!\!\!\!\!\!W^{\na}(y^{\na}|x^{*{\na}}) &= \sum_{y^{\na} \in \mathcal{D}^*}W^{\na}(y^{\na}|x^{*{\na}}) \psi_{y^{\na}}
    \end{align}
    and with Lemma \ref{lem:weighted-chernoff-bernoulli} from the Appendix\footnote{In \cite{mross_second-order_2025}, we mistakenly claimed that Bernstein's inequality can be used.}, we find that, with $1/M'' < \mu_{\na} <1$,
    \begin{align}
    &\Pr[\bar{\lambda}^{(i_1, i_2)}_2 > \mu_{\na} + \mu_{\nb}'']\\
        &\leq\Pr\left[\sum_{y^{\na} \in \mathcal{D}^*}W^{\na}(y^{\na}|x^{*{\na}}) \psi_{y^{\na}} >  \mu_{\na}\right] \nonumber\\
        &\leq \exp\left({-\rho_{\na}D\left(\mu_{\na}\middle\|\frac{1}{M''}\right) }\right). \\
        &\leq \exp\left({-\rho_{\na}\cdot (\mu_{\na} {\nb} R_c'' -1)}\right)
    \end{align}
    where in the last step, similarly to \cite{ahlswede_identification_1989-1}, we use $M'' = \left\lceil2^{{\nb} R_c''}\right\rceil$ to bound
    \begin{align}
        D\left(\mu_{\na}\middle\|\frac{1}{M''}\right) \geq \mu_{\na} {\nb} R_c'' -1.\label{eq:divergence-lambda-m}
    \end{align}
    
    The constructed code is not an $(\ntot, N, \lambda_1, \lambda_2)$-IDf code if the condition on the error of the second kind is {not fulfilled} for any combination of $i_1, i_2$, i.e.,\footnote{This argument is not fully present in \cite{ahlswede_identification_1989-1}, but can be found e.g. in \cite[Proof of Theorem 40]{ahlswede_identification_2021}}
    \begin{align}
        &\Pr\left[\mathcal{C} \text{ is not an } (\ntot, N, \lambda_1, \lambda_2)\text{-IDf code}\right] \nonumber\\
        &= \Pr\left[\bigcup_{\substack{i_1, i_2 \in \{1, ... N\} \\ i_1 \neq i_2}}\bar{\lambda}^{(i_1, i_2)}_2 > \mu_{\na}  + \mu_{\nb}'' \right] \nonumber\\
        &\leq \sum_{i_1 = 1}^{N} \sum_{\substack{i_2 = 1 \\ i_2 \neq i_1}}^{N} \Pr[\bar{\lambda}^{(i_1, i_2)}_2 > \mu_{\na} + \mu_{\nb}''] \nonumber\\
        &\leq N^2 \exp\left({-\rho_{\na} \cdot (\mu_{\na} {\nb} R_c'' -1) }\right).
    \end{align}
    By the random coding argument, we know that at least one code exists if this probability is strictly smaller than one, i.e.,
    \begin{align}
        N &< {\exp\left(\frac{1}{2}\rho_{\na} \cdot (\mu_{\na} {\nb} R_c''-1) \right)}. \label{eq:idf-ach2-n-bound}
    \end{align}
    Now choose an $N$ fulfilling \eqref{eq:idf-ach2-n-bound}, e.g. 
    \begin{align}
        N &= \left\lceil{\exp\left(\frac{1}{2}\rho_{\na} \cdot (\mu_{\na} {\nb} R_c''-1) \right)}\right\rceil-1.
    \end{align}
    This results in
    \begin{align}
        \log \log N =& {\na}H(W(\cdot | x^{*}))- \sqrt{{\na}V_{\mathsf{IDf}}(\lambda_1, W)}Q^{-1}(\mu_{\na}') \nonumber\\ &+ O(\log {\na}).\label{eq:idf-ach2-loglogn-bound}
    \end{align}    
    First, we normalize \eqref{eq:idf-ach2-loglogn-bound} w.r.t. the correct blocklength ${\ntot}$. Note that $\frac{{\na}}{{\ntot}} = 1 + O\left(\frac{\log {\na}}{{\na}}\right)$ and $\frac{\sqrt{{\na}}}{{\ntot}} 
    = \sqrt{\frac{1}{{\ntot}}} +O\left( \frac{\log {\na}}{{\na}^{3/2}}\right)$. Thus, using $O\left(\frac{\log {\na}}{{\na}}\right) = O\left(\frac{\log {\ntot}}{{\ntot}}\right)$,
\begin{align}
        \frac{ \log \log N}{{\ntot}} =& H(W(\cdot | x^{*})) - \sqrt{\frac{V_{\mathsf{IDf}}(\lambda_1, W)}{{\ntot}}}Q^{-1}(\mu_{\na}') \nonumber\\&+ O\left(\frac{\log {\ntot}}{\ntot}\right). \label{eq:idf-ach2-completion}
    \end{align}   
    
We have now established that
    \begin{align}
        \lambda_1^{(i)} &\leq \mu_{\na}' + \delta_{\na} + \mu_{\nb}'' \leq \lambda_1\\
        \lambda_2^{(i_1, i_2)} &\leq \mu_{\na} + \mu_{\nb}'' \leq \lambda_2,
    \end{align}
    where $\lambda_1$ and $\lambda_2$ are the constraints on the error of the first and second kind, respectively. We can choose the code rate $R_c'' < C(W)$, such that we operate in the large deviation regime and \begin{align}
            \mu_{\nb}'' &\leq \exp(-\gamma\log {\ntot}E(R_c'') + o(\log {\ntot})) = O\left(\frac{1}{\ntot}\right){,}
        \end{align}
        where $E(R)$ is the error exponent and we choose $\gamma$ such that $\gamma E(R_c'') > 1$. By the Berry-Esseen Theorem, we have $\delta_{\na} = O(1/\sqrt{\na}) = O(1/\sqrt{\ntot})$. For the choice of $\mu_{\na}$, note that we can set $\mu_{\na} = \lambda_2 - \mu_{\nb}''$. Then we can set $\mu_{\na}' = \lambda_1 - \delta_{\na} - \mu_{\nb}'' = \lambda_1 + O\left(\frac{1}{\sqrt{\na}}\right)$, and thus, we have
        \begin{align}
            Q^{-1}(\mu_{\na}') &= Q^{-1}(\lambda_1) + O\left(\frac{1}{\sqrt{\ntot}}\right)
        \end{align}
    Inserting this into \eqref{eq:idf-ach2-completion}, we get
    \begin{align}
        &\frac{\log \log N_{\mathsf{IDf}}^*({\ntot}, \lambda_1, \lambda_2|W)}{\ntot} \geq \frac{ \log \log N}{{\ntot}} \\
        &= H(W(\cdot | x^{*})) \!- \!\sqrt{\frac{V_{\mathsf{IDf}}(\lambda_1, W)}{\ntot}}Q^{-1}(\lambda_1) \!+\! O\left(\frac{\log {\ntot}}{\ntot}\right). \label{eq:idf-ach2-completion2}
    \end{align}   
    Taking the limits according to Def. \ref{def:idf-det-capacity} and \ref{def:idf-det-second-order}, we obtain the desired result.

\subsection{Achievability (Zero variance)}
Assume $V_{\mathsf{IDf}}(\lambda_1, W) = 0$. This implies that 
\begin{align}
    \log \frac{1}{W^{\na}(Y^{\na}|x^{*\na})} = \na H(W(\cdot|x^*))
\end{align}
almost surely, which in turn implies
$W^{\na}(\mathcal{D}^*|x^{*\na}) = 1$
when using the definition of the set $\mathcal{D}^*$ in \eqref{eq:idf-ach2-set-d} and $\log \rho_{\na} = \na H(W(\cdot|x^*))$. Continuing the same construction as in the nonzero variance case, this implies that $\lambda_1^{(i)} \leq \mu_{\nb}''$.
The analysis of the second-kind error proceeds as in the nonzero variance case. Thus, we obtain
\begin{align}
        &\frac{\log \log N_{\mathsf{IDf}}^*({\ntot}, \lambda_1, \lambda_2|W)}{\ntot} \geq  H(W(\cdot | x^{*})) + O\left(\frac{\log {\ntot}}{\ntot}\right). 
    \end{align}   
    
\subsection{Converse}\label{sec:idf-det-converse}
For the converse proof, we use the following Lemma\footnote{In \cite{mross_second-order_2025}, we mistakenly stated this Lemma with a $\max$ instead of a $\min$, and the $\min$ and $\max$ in Lemma \ref{lem:idf-rand-converse-lemma} were mixed up.}, which is an extension of an argument in \cite{ahlswede_identification_1989-1} to the case $\lambda_1 + \lambda_2 < 1$.

\begin{lemma}\label{lem:idf-det-converse-lemma}
    For $\rho_n > 0$, let
    \begin{equation}
        \mathcal{E}_{\vec{f}} = \Big\{y^n\in\mathcal{Y}^n:\log \frac{1}{W^n(y^n|{\vec{f}})} \leq \log \rho_n \Big\}. \label{eq:id-feedback-converse-lemma-set}
    \end{equation}
    Suppose that
    \begin{align}
         \min_{\vec{f}} W^n(\mathcal{E}_{\vec{f}}|{\vec{f}}) \geq 1-\nu \label{eq:id-feedback-converse-lemma-condition}
    \end{align}
    for some $\nu > 0$. Then any $(n, N, \lambda_1, \lambda_2)$-IDf code with $\lambda_1+\lambda_2 < 1 - \nu$ satisfies
    \begin{align*}
        \log \log N^*(n, \lambda_1, \lambda_2) \leq \log \rho_n + \log n + \log \log |\mathcal{Y}|.
    \end{align*}
\end{lemma}
    \begin{IEEEproof}
        For any $i \in [N]$, note that
    \begin{align}
        W^n(\mathcal{D}_i \cap \mathcal{E}_{{\vec{f}}_i}|{\vec{f}}_i) &= 1- W^n(\mathcal{D}_i^c \cup \mathcal{E}_{{\vec{f}}_i}^c|{\vec{f}}_i)\\
        &\geq 1- W^n(\mathcal{D}_i^c|{\vec{f}}_i) - W^n(\mathcal{E}_{f_i}^c|{\vec{f}}_i) \\
        &\geq 1-\lambda_1-\nu, \label{eq:id-feedback-converse-lemma-proof-intersection-probability-bound}
    \end{align}
    where the last step follows from the condition on the error of the first kind and the assumption \eqref{eq:id-feedback-converse-lemma-condition}. For the sake of contradiction, assume that there exists a $j \in \{1, ..., N\}, j\neq i$, such that $\mathcal{D}_j \cap \mathcal{E}_{{\vec{f}}_j} = \mathcal{D}_i \cap \mathcal{E}_{{\vec{f}}_i}$. Then by \eqref{eq:id-feedback-converse-lemma-proof-intersection-probability-bound} we find that
    \begin{align}
         1-\lambda_1-\nu &\leq W^n(\mathcal{D}_j \cap \mathcal{E}_{{\vec{f}}_j}|{\vec{f}}_i) \\
        &\leq W^n(\mathcal{D}_j|{\vec{f}}_i) \\
        &\leq \lambda_2,
    \end{align}
    due to the bound on the second-kind error, which contradicts the assumption that $\lambda_1 + \lambda_2 < 1-\nu$. Therefore, the sets $\mathcal{D}_i \cap \mathcal{E}_{{\vec{f}}_i}$ must be distinct.
    
    This means that the total number of messages is upper bounded by the number of such intersection sets. By definition, $\mathcal{E}_{{\vec{f}}}$ contains all vectors $y^n$ satisfying $W^n(y^n|{\vec{f}}) \geq \frac{1}{\rho_n},$
    which implies $|\mathcal{E}_{\vec{f}}| \leq \rho_n$. Therefore, we also have $|\mathcal{D}_i \cap \mathcal{E}_{{\vec{f}}_i}| \leq \rho_n$, i.e., $N$ is upper bounded by the number of subsets of cardinality $\leq \rho_n$:
    \begin{align}
        N \leq \sum_{k=1}^{\min\{\rho_n, |\mathcal{Y}|^n\}} {\binom{|\mathcal{Y}|^n}{k}} \leq 2^{n\log |\mathcal{Y}| \cdot \rho_n}.
    \end{align}
    \end{IEEEproof}
The first-order converse for \eqref{eq:idf-det-capacity} can now be shown in the same way as in \cite{ahlswede_identification_1989-1} using the Chebychev inequality.

Now we proceed to the second-order converse and assume that $W$ is IDf-simple dispersion. Due to the feedback, the sequence
\begin{align}
    \log \frac{1}{W(Y_t|{\vec{f}}(Y^{t-1}))} - \expect{\log \frac{1}{W(Y_t|{\vec{f}}(Y^{t-1}))}|Y^{t-1}}
\end{align}
cannot be assumed to be i.i.d. and thus, we use a central limit theorem for martingale difference sequences \cite{bolthausen_exact_1982} and the Azuma-Hoeffding inequality to analyze the probability \eqref{eq:id-feedback-converse-lemma-condition}. 

To analyze the probability expression in Lemma \ref{lem:idf-det-converse-lemma}, we use the technique from \cite[Theorem 3]{wagner_new_2020} and adapt it to the set \eqref{eq:id-feedback-converse-lemma-set} and deterministic feedback encoding functions ${\vec{f}}$.
Define
    \begin{align}
        \nu_{\mathsf{IDf}}^{x}(W) &:= \Var\left[\log \frac{1}{W(Y|x)}\right],\\
        \nu_{\mathsf{IDf}}^{\max}(W) &:= \max_{x \in \mathcal{X}} \nu_{\mathsf{IDf}}^{x}(W).
    \end{align}
Let the set of optimal input symbols $\mathcal{X}^*_W$ be given by \eqref{eq:optimal-input-set-det}. 
    For a sequence $x^n \in \mathcal{X}^n$, define
    \begin{align}
        \phi_W(x^n) := \sum_{x \notin \mathcal{X}^*_W} P_{x^n}(x),
    \end{align}
    which is a measure for the ratio of non-optimal input symbols in the vector $x^n$.
    Furthermore, fix some $\gamma > 0$, 
    and define the set 
    \begin{align}
        \mathcal{A}^{\gamma}_{\vec{f}} := \left\{y^n \in \mathcal{Y}^n: \phi_W({\vec{f}}(Y^n)) \leq \gamma\right\}.
    \end{align}
    Nox fix a $x^* \in \mathcal{X}^*_W$ and let $\vec{f}_* = (f^1_*, ..., f^n_*)$ be defined by
    \begin{align}
        f_{*}^t(y^{t-1}) = \begin{cases}
            f^t(y^{t-1}) &\text{if } \sum_{\tau = 1}^t \indic\left\{f^{\tau}(y^{\tau-1}) \notin \mathcal{X}^*_W\right\} \leq \lfloor n\gamma\rfloor \\
            x^* & \text{otherwise.}
        \end{cases}
    \end{align}
    With this definition, we have $W^n(y^n|\vec{f}) = W^n(y^n|\vec{f}_*)$ for all $y^n \in  \mathcal{A}^{\gamma}_{\vec{f}}$. For brevity, define $\Pr_*[\cdot] = W^n(\cdot|\vec{f}_*)$ and $\expect[*]{\cdot}$ and $\Var_*[\cdot]$ defined accordingly.

We can write
\begin{align}
    W^n(\mathcal{E}_{\vec{f}}|f) &= 1- W^n(\mathcal{E}_{\vec{f}}^c|f) \\
    &= 1- \Pr\left[\sum_{t=1}^n \log \frac{1}{W(Y_t|f(Y^{t-1}))} > \log \rho_n  \right].
\end{align}
Let $\mathcal{F}_t = \sigma(Y_1, ..., Y_t)$ and $X_t^* = f_*^t(Y^{t-1})$. Note that the sequence 
\begin{align}
    \tilde{Z}_t := \log \frac{1}{W(Y_t|X_t^*)} - \expect[*]{\log \frac{1}{W(Y_t|X_t^*)}\middle|Y^{t-1}}
\end{align}
forms a martingale difference sequence and consequently, the sequence $Z_n:=\sum_{t=1}^n \tilde{Z}_t$
forms a martingale. We can write
\begin{align}
    W^n(\mathcal{E}_{\vec{f}}^c|\vec{f}) &= W^n(\mathcal{E}_{\vec{f}}^c \cap \mathcal{A}^{\gamma}_{\vec{f}}|\vec{f}) + W^n(\mathcal{E}_{\vec{f}}^c \cap \mathcal{A}^{\gamma, c}_{\vec{f}}|\vec{f})\\
    &= W^n(\mathcal{E}_{\vec{f}_*}^c \cap \mathcal{A}^{\gamma}_{\vec{f}}|\vec{f}_*) + W^n(\mathcal{E}_{\vec{f}}^c \cap \mathcal{A}^{\gamma, c}_{\vec{f}}|\vec{f})\\
    &\leq W^n(\mathcal{E}_{\vec{f}_*}^c |\vec{f}_*) + W^n(\mathcal{E}_{\vec{f}}^c \cap \mathcal{A}^{\gamma, c}_{\vec{f}}|\vec{f})\label{eq:idf-det-converse-prob-bound}
\end{align}
We start by analyzing the first probability in \eqref{eq:idf-det-converse-prob-bound}. 
For all $y^{t-1}\in\mathcal{Y}^{t-1}$, we have
\begin{align}
    &\expect[*]{\log\frac{1}{ W(Y_t|{f^t_*}(Y^{t-1}))}\middle| Y^{t-1} = y^{t-1}} \nonumber\\
    &= \sum\limits_{y_t \in\mathcal{Y}} \log \frac{1}{W(y_t|f^t_{*}(y^{t-1}))} \cdot W(y_t|f^t_{*}(y^{t-1})) \notag \\
    &= H(W(\cdot|f^t_{*}(y^{t-1})) \notag \\
    &\leq H(W(\cdot|x^*)),
\end{align}
where $x^* \in \mathcal{X}^*$, and therefore
\begin{align}
    \expect[*]{\log\frac{1}{ W(Y_t|f_*^t(Y^{t-1}))}\middle|Y^{t-1}}
    &\leq H(W(\cdot|x^*)).
\end{align}

\begin{lemma}\label{lem:idf-det-converse-lemma-closetypes}
    Assume that $V_{\mathsf{IDf}}^{\min}(W) = V_{\mathsf{IDf}}^{\max}(W) =V_{\mathsf{IDf}}(W) > 0$. 
    Then with $K_W = \max\left\{\nu_{\mathsf{IDf}}^{\max}(W), \frac{4\nu_{\mathsf{IDf}}^{\max}(W)}{V_{\mathsf{IDf}}(W)}\right\}$ and $\gamma \in \left(0, \frac{V_{\mathsf{IDf}}(W)}{2K_W}\right)$, we have 
    \begin{align}
       V_{\mathsf{IDf}}(W)-\gamma K_W \leq \frac{1}{n}\sum_{t=1}^n \expect[*]{\tilde{Z}_t^2|\mathcal{F}_{t-1}} &\leq V_{\mathsf{IDf}}(W) + \gamma K_W,\\
        \left\|\frac{\sum_{t=1}^n \expect[*]{\tilde{Z}_t^2|\mathcal{F}_{t-1}}}{\sum_{t=1}^n \expect[*]{\tilde{Z}_t^2}}-1\right\|_{\infty} &\leq \gamma K_W.
    \end{align}
\end{lemma}
\begin{IEEEproof} 
By construction of $\vec{f}_*$, we have 
\begin{align}
    \phi_W(\vec{f}_*(Y^n)) \leq \gamma.
\end{align}
Thus, it follows that
\begin{align}
    &\frac{1}{n}\sum_{t=1}^n \expect[*]{\tilde{Z}_t^2|\mathcal{F}_{t-1}} \\
    &= \frac{1}{n}\sum_{t=1}^n\Var_*\left[\log \frac{1}{W(Y_t|X^*_t)}\middle|Y^{t-1}\right] \\
    &= \frac{1}{n}\sum_{t=1}^n\sum_{x\in \mathcal{X}}\indic[X_t^* = x]\Var_*\left[\log \frac{1}{W(Y_t|X)}\middle|X=x\right] \\
    &= \sum_{x\in \mathcal{X}} P_{X^{*n}}(x) \nu_{\mathsf{IDf}}^{x}(W) \\
    &= \sum_{x\in \mathcal{X}^*} P_{X^{*n}}(x) \nu_{\mathsf{IDf}}^{x}(W) + \sum_{x\notin \mathcal{X}^*} P_{X^{*n}}(x) \nu_{\mathsf{IDf}}^{x}(W)\\
    &\leq V_{\mathsf{IDf}}(W) + \gamma \nu_{\mathsf{IDf}}^{\max}(W).
\end{align}
Similarly, we obtain the lower bound 
\begin{align}
    \frac{1}{n}\sum_{t=1}^n \expect[*]{\tilde{Z}_t^2|\mathcal{F}_{t-1}}\geq& V_{\mathsf{IDf}}(W) - \gamma \nu_{\mathsf{IDf}}^{\max}(W)
\end{align}

For the second assertion, note that 
\begin{align}
    \left|\frac{\sum_{t=1}^n \expect[*]{\tilde{Z}_t^2|\mathcal{F}_{t-1}}}{\sum_{t=1}^n \expect[*]{\tilde{Z}_t^2}}\!-\!1\right| &\!\leq\! \left|\frac{V_{\mathsf{IDf}}(W) + \gamma \nu_{\mathsf{IDf}}^{\max}(W)}{V_{\mathsf{IDf}}(W) - \gamma \nu_{\mathsf{IDf}}^{\max}(W)}\!-\!1\right| \\
    &= \frac{2\gamma \nu_{\mathsf{IDf}}^{\max}(W)}{V_{\mathsf{IDf}}(W)-\gamma\nu_{\mathsf{IDf}}^{\max}(W)}. \\
    &\leq \frac{2\gamma \nu_{\mathsf{IDf}}^{\max}(W)}{V_{\mathsf{IDf}}(W)-\frac{1}{2}V_{\mathsf{IDf}}(W)}. \\
    &\leq \frac{4\gamma\nu_{\mathsf{IDf}}^{\max}(W)}{V_{\mathsf{IDf}}(W)},
\end{align}
where we used $\gamma \leq \frac{V_{\mathsf{IDf}}(W)}{2\nu_{\mathsf{IDf}}^{\max}(W)}$.
\end{IEEEproof}

We abbreviate $H(W(\cdot|x^*)) =: H$ and let
\begin{align}
    V_{\gamma}(u) := \begin{cases}
       V_{\mathsf{IDf}}(W)-\gamma K_W &0 < u \leq \frac{1}{2} , \\
        V_{\mathsf{IDf}}(W)+\gamma K_W, &\frac{1}{2} < u < 1.
    \end{cases}
\end{align}
and choose
\begin{align}
    \log \rho_n = nH - \sqrt{nV_{\gamma}(\lambda_n)}Q^{-1} (\lambda_n).
\end{align}
where $\lambda_n$ is to be specified later.
Then we can bound
\begin{IEEEeqnarray}{rCl}
\IEEEeqnarraymulticol{3}{l}{{\Pr}_*\left[\sum_{t=1}^n \log \frac{1}{W(Y_t|f^t_*(Y^{t-1}))} > \log \rho_n \right]} \\
&=&{\Pr}_*\left[\frac{\sum_{t=1}^n \log \frac{1}{W(Y_t|f_*^t(Y^{t-1}))} - nH}{\sqrt{nV_{\gamma}(\lambda_n)}} > -Q^{-1} (\lambda_n)\right]  \nonumber\\\\
&\leq& {\Pr}_*\left[\frac{\sum_{t=1}^n \tilde{Z}_t}{\sqrt{\sum_{t=1}^n\expect[*]{\tilde{Z}_t^2}}} > -Q^{-1} (\lambda_n)\right]\nonumber\\\label{eq:idf-det-converse-closetypes-lemma-application} \\
&\leq&Q\left(-Q^{-1} (\lambda_n)\right) 
+ L\Bigg(\frac{n \log n}{\left(\sum_{t=1}^n\expect[*]{\tilde{Z}_t^2}\right)^{3/2}} \nonumber\\*
&&+ \left\|\frac{\sum_{t=1}^n \expect[*]{\tilde{Z}_t^2|\mathcal{F}_{t-1}}}{\sum_{t=1}^n \expect[*]{\tilde{Z}_t^2}} -1\right\|_{\infty}^{1/2}\Bigg)\label{eq:idf-det-converse-bet-application}\\
&\leq& 1- \lambda_n + \underbrace{L \left(\frac{\log n}{\sqrt{n} (V_{\mathsf{IDf}}(W)-\gamma K_W)^{3/2}} + \sqrt{\gamma K_W}\right)}_{=:\delta_{1,n}} \nonumber\\\label{eq:idf-det-converse-bet-simplification}\\
&=& 1-\lambda_n + \delta_{1,n}, \label{eq:idf-det-converse-closetypes-bound}
\end{IEEEeqnarray}
where \eqref{eq:idf-det-converse-closetypes-lemma-application} follows from Lemma \ref{lem:idf-det-converse-lemma-closetypes},  \eqref{eq:idf-det-converse-bet-application} follows from the Martingale Berry-Esseen Thm. \cite[Corollary to Theorem 2]{bolthausen_exact_1982}, and \eqref{eq:idf-det-converse-bet-simplification} follows from $Q(-x) = 1-Q(x)$ and Lemma \ref{lem:idf-det-converse-lemma-closetypes}.

Now we consider the second term in \eqref{eq:idf-det-converse-prob-bound}. Let $X_t := f^t(Y^{t-1})$ and define
\begin{align}
    \mathcal{T}_{\gamma}^n := \left\{T \in \mathcal{P}_n(\mathcal{X}): \sum_{x \notin \mathcal{X}_W^*} T(x) > \gamma \right\}
\end{align}
If $\mathcal{X}_W^* = \mathcal{X}$, then $(\mathcal{A}_f^{\gamma})^c$ is empty and we take $\delta_{2,n} =0$. Otherwise, fix $T \in \mathcal{T}_{\gamma}^n$. On the event $\{P_{X^n} = T\}$, we have
\begin{align}
    &\sum_{t=1}^n \expect{\log \frac{1}{W(Y_t|X_t)} \middle| Y^{t-1}} \\
    &= \sum_{x\in\mathcal{X}}\sum_{t=1}^n H(W(\cdot|x)) \indic[f(Y^{t-1}) =x]\\
    &= n\sum_{x\in\mathcal{X}}T(x)H(W(\cdot|x))\\
    &= nH(W|T).
\end{align}
Abbreviate $H(W(\cdot|x^*))=:H$ and define
\begin{align}
    \Delta_W := \min_{x\notin \mathcal{X}_W^*}[H-H(W(\cdot|x))] > 0.
\end{align}
For every $T \in \mathcal{T}_{\gamma}^n$, this gives
\begin{align}
    K_T &:= H-H(W(\cdot|x)) \\
    &= \sum_{x \notin \mathcal{X}^*} T(x) [H-H(W(\cdot|x))] \geq \gamma \Delta_W.
\end{align}

Let $Z_0^{\vec{f}} = 0$ and
\begin{align}
    Z_t^{\vec{f}} := \sum_{\tau = 1}^t \left(\log \frac{1}{W(Y_{\tau}|X_{\tau})} - H(W(\cdot|X_{\tau}))\right).
\end{align}
The sequence $(Z_t^{\vec{f}})_{t=0}^n$ is a martingale. 
Let $W_{\min} := \min_{x,y: W(y|x)>0} W(y|x)$. We now bound the martingale differences:
\begin{align}
    |Z_t^{\vec{f}} - Z_{t-1}^{\vec{f}}| &= \left|\log \frac{1}{W(Y_t|X_t)}  - H(W(\cdot|X_t)) \right| \\
    &\leq 2\log  \max_{x,y: W(y|x) > 0} \frac{1}{W(y|x)} \\
    &\leq 2\log \frac{1}{W_{\min}}.
\end{align}
We can now bound, for each $T \in \mathcal{T}_{\gamma}^n$,
\begin{align}
    &{\Pr}\left[\sum_{t=1}^n \log \frac{1}{W(Y_t|X_t)} > \log \rho_n , \; P_{X^n} = T\right] \\
    &= {\Pr}\left[Z_n^{\vec{f}} > \log \rho_n - nH(W|T), \; P_{X^n} = T\right] \\
    &\leq {\Pr}\left[Z_n^{\vec{f}} > nK_T- \sqrt{nV_{\gamma}(\lambda_n)}Q^{-1} (\lambda_n)\right] \\
    &\leq {\Pr}\left[Z_n^{\vec{f}} > \frac{nK_T}{2}\right] \label{eq:idf-det-converse-fartypes-nlarge}\\
    &\leq \exp\left(-\frac{\frac{n^2K_T^2}{4}}{2\cdot n\cdot 4\log^2 \frac{1}{W_{\min}}}\right) \label{eq:idf-det-converse-azuma-hoeffding}\\
    &=\exp\left(-\frac{nK_T^2}{32 \log^2 \frac{1}{W_{\min}}}\right)\\
    &\leq \exp\left(-nK\right),
\end{align}
where \eqref{eq:idf-det-converse-fartypes-nlarge} holds uniformly over all $T \in \mathcal{T}^n_\gamma$ for sufficiently large $n$ and we define $K:= \frac{\gamma^2\Delta_W^2}{32 \log^2 \frac{1}{W_{\min}}} > 0$. In \eqref{eq:idf-det-converse-azuma-hoeffding}, we used the Azuma-Hoeffding inequality \cite[Theorem 7.2.1]{alon_probabilistic_2008}.
Summing over all possible type classes, we have 
\begin{align}
&W^n(\mathcal{E}_{\vec{f}}^c \cap \mathcal{A}^{\gamma, c}_{\vec{f}}|\vec{f})\\
    &=\sum_{T \in \mathcal{T}_{\gamma}^n} {\Pr}\left[\sum_{t=1}^n \log \frac{1}{W(Y_t|X_t)} > \log \rho_n , \; P_{X^n} = T\right]\\
    &\leq (n+1)^{|\mathcal{X}|} \exp\left(-nK\right) \\
    &=: \delta_{2,n}. \label{eq:idf-det-converse-fartypes-bound}
\end{align}
Inserting \eqref{eq:idf-det-converse-closetypes-bound} and \eqref{eq:idf-det-converse-fartypes-bound} into \eqref{eq:idf-det-converse-prob-bound}, we obtain
\begin{align}
    W^n(\mathcal{E}_{\vec{f}}^c|{\vec{f}}) &\leq 1-\lambda_n+\delta_{1,n}+\delta_{2,n}\\
    \Leftrightarrow W^n(\mathcal{E}_{\vec{f}}|{\vec{f}}) &\geq \lambda_n-\delta_{1,n}-\delta_{2,n}.\label{eq:idf-det-converse-prefinal-error-bound}
\end{align}

Choose  $\lambda_n=\lambda_1+\lambda_2+2(\delta_{1,n}+\delta_{2,n})$. For sufficiently small $\gamma >0$ and large $n$, we have $\lambda_n < 1$. By \eqref{eq:idf-det-converse-prefinal-error-bound}, we have
\begin{align}
    \inf_{\vec{f}} W^n(\mathcal E_{\vec{f}}\mid \vec{f})
&\geq \lambda_n-\delta_{1,n}-\delta_{2,n} \\
&= \lambda_1 + \lambda_2 + \delta_{1,n} + \delta_{2,n} \\
&\geq \lambda_1 + \lambda_2.
\end{align}
Lemma \ref{lem:idf-det-converse-lemma} therefore yields
\begin{align}
    \log\log N_{\mathrm{IDf}}^*(n,\lambda_1,\lambda_2\mid W)
\leq& nH-\sqrt{nV_\gamma(\lambda_n)}Q^{-1}(\lambda_n) \nonumber\\
&+O(\log n).
\end{align}
For fixed $\gamma$, the error term yields
\begin{align}
    \lim_{n \to \infty} \lambda_n = \lambda_1 + \lambda_2 + 2L\sqrt{\gamma K_W}.
\end{align}
Taking $\gamma\to 0$ and $\lambda_2 \to 0$ proves the desired second-order converse.

\section{Proof of Theorem \ref{thm:second-order-idf-rand}}\label{sec:proof-rand}
\subsection{Uniqueness of the Dispersion: Proof of Theorem \ref{thm:idf-rand-simple-dispersion}}\label{sec:proof-unique-dispersion}
We need the following lemma, which extends a classic result \cite[Corollary 2 to Theorem 4.5.2.]{gallager_information_1972} to the IDF setting.
\begin{lemma}\label{lem:idf-rand-unique-output}
    The capacity-achieving output distribution of randomized identification with feedback is unique.
\end{lemma}
\begin{IEEEproof}
    We use the fact that the entropy $H(Q)$ is strictly concave in $Q$, i.e., for two distinct $Q_1, Q_2 \in \mathcal{P}(\mathcal{Y})$ and $\theta \in (0,1)$, we have
    \begin{align}
        H(\theta Q_1 + (1-\theta)Q_2) > \theta H(Q_1) + (1-\theta) H(Q_2).
    \end{align}
    Now assume that there exist two distinct $Q_1, Q_2 \in \mathcal{P}(\mathcal{Y})$ that both achieve capacity $H(Q_1) = H(Q_2) = C$, and fix some $\theta \in (0,1)$. By strict concavity, we have
    \begin{align}
        H(\theta Q_1 + (1-\theta)Q_2) &> \theta H(Q_1) + (1-\theta) H(Q_2)= C,
    \end{align}
    which contradicts the assumption that $C$ is the capacity. Thus, the capacity-achieving output distribution is unique.
\end{IEEEproof}

The proof of Theorem \ref{thm:idf-rand-simple-dispersion} follows from the fact that
    the IDF-dispersion $V_{\mathsf{IDF}}^{\max}(W) = \max_{P \in  \Phi_W^*}\Var\left[\log \frac{1}{P W(Y)}\right]$ depends only on the output distribution $PW$. By Lemma \ref{lem:idf-rand-unique-output}, the capacity-achieving output distribution is unique, and thus, there is only one possible value for the dispersion.

\subsection{Achievability}\label{sec:idf-rand-achievability}
The proof follows the same steps as the achievability proof of Theorem \ref{thm:second-order-idf}. The transmitter sends a random sequence $X^{\na}$ i.i.d. generated according to a capacity-achieving input distribution $P^*$ instead of transmitting $x^{*\na}$. Furthermore, the set $\mathcal{D}^*$ is replaced by
\begin{align}
    \mathcal{D}^{**} &= \left\{y^{\na} \in \mathcal{Y}^{\na}: \, \log \frac{1}{P^*W^{\na}(y^{\na})} \geq \log \rho_{\na}\right\}, \\
    \log \rho_{\na} &= \na H(P^*W) -\sqrt{\na V_{\mathsf{IDF}}(W)}Q^{-1}(\mu_{\na}').
\end{align}

\subsection{Converse}\label{sec:idf-rand-converse}
For the converse, we adapt Lemma \ref{lem:idf-det-converse-lemma} to the case where randomized channel inputs are possible.
\begin{lemma}\label{lem:idf-rand-converse-lemma}
    For $\rho_n > 0$, let
    \begin{equation}
        \mathcal{E}_{P_{Y^n}} = \Big\{y^n\in\mathcal{Y}^n:\log \frac{1}{P_{Y^n}(y^n)} \leq \log \rho_n \Big\}.\label{eq:id-feedback-rand-converse-lemma-set}
    \end{equation}
    Suppose that
    \begin{align}
         \max_{P_{Y^n}\in \mathcal{P}(\mathcal{Y}^n)}\min_{Q_F\in \mathcal{P}(\mathcal{F}_n)} Q_F W^n(\mathcal{E}_{P_{Y^n}}) \geq 1-\nu \label{eq:id-feedback-rand-converse-lemma-condition}
    \end{align}
    for some $\nu > 0$. Then any $(n, N, \lambda_1, \lambda_2)$-IDF code with $\lambda_1+\lambda_2 < 1 - \nu$ satisfies
    \begin{align*}
        \log \log N^*(n, \lambda_1, \lambda_2) \leq \log \rho_n + \log n + \log \log |\mathcal{Y}|.
    \end{align*}
\end{lemma}
\begin{IEEEproof}
    For brevity, we write $Q_i := Q_F(\cdot|i)$ for the randomized feedback encoding functions. For any $i \in [N]$, note that
    \begin{align}
        Q_iW^n(\mathcal{D}_i \cap \mathcal{E}_{P_{Y^n}}) &= 1- Q_iW^n(\mathcal{D}_i^c \cup \mathcal{E}_{P_{Y^n}}^c)\\
        &\geq 1- Q_iW^n(\mathcal{D}_i^c) - Q_iW^n(\mathcal{E}_{P_{Y^n}}^c) \\
        &\geq 1-\lambda_1-\nu \label{eq:id-feedback-rand-converse-lemma-proof-intersection-probability-bound}
    \end{align}
    where the last step follows from the condition on the error of the first kind and the assumption \eqref{eq:id-feedback-converse-lemma-condition}. Note that, to fulfill the error requirements, $\mathcal{D}_i \cap \mathcal{E}_{P_Y}$ must be distinct, so the total number of messages is upper bounded by the number of such intersection sets. By definition, $\mathcal{E}_{P_{Y^n}}$ contains all vectors $y^n$ satisfying $
        P_{Y^n}(y^n) \geq \frac{1}{\rho_n},$
    which implies $|\mathcal{E}_{P_{Y^n}}|< \rho_n$. Therefore, we also have $|\mathcal{D}_i \cap \mathcal{E}_{P_{Y^n}}| \leq \rho_n$, i.e., $N$ is upper bounded by the number of subsets of cardinality $\leq \rho_n$:
    \begin{align}
        N \leq \sum_{k=0}^{\rho_n} {\binom{|\mathcal{Y}|^n}{k}} \leq 2^{n\log |\mathcal{Y}| \cdot \rho_n}.
    \end{align}
\end{IEEEproof}
Let $P_Y^* = P^*W$ be the unique capacity-achieving output distribution. We disregard output symbols with $W(y|x) = 0$ for all $x$, such that we can assume $P_Y^*(y) > 0$ got all remaining outputs. Define $\mathcal{F}_t = \sigma(U, Y^t)$, where $U$ captures the local randomness at the encoder.  

First, we consider the case $V_{\mathsf{IDF}}=0$. Choose $P_{Y^n}=(P_Y^*)^n$ and $\rho_n = |\mathcal{Y}|^n$. Then $\mathcal{E}_{P_{Y^n}} = \mathcal{Y}^n$. Now Lemma \ref{lem:idf-rand-converse-lemma} applies and directly gives the desired converse bound.

We now consider the case $V_{\mathsf{IDF}}>0$. Let the set of optimal input distributions $\Phi^*_W$ be given by \eqref{eq:optimal-input-set-rand}. 
    For a sequence $x^n \in \mathcal{X}^n$, define
    \begin{align}
        \psi_W(x^n) := \inf_{P \in \Phi_W^*} d_{\text{TV}}(P, P_{x^n}),
    \end{align}
    which is a measure for the distance of the vector $x^n$ to an optimal input type. Here, $d_{\text{TV}}$ denotes the total variation distance.
    Furthermore, fix some $\gamma >0$ and sufficiently small, as specified later. Define the set
    \begin{align}
        \mathcal{S}_{\gamma}^{n} := \left\{T \in \mathcal{P}_n(\mathcal{X}): \inf_{P \in \Phi_W^*} d_{\text{TV}}(P,T) > \gamma\right\}.
    \end{align}
    Again, the sequence 
\begin{align}
    \tilde{Z}'_t := \log \frac{1}{P_Y^*(Y_t)} - \expect{\log \frac{1}{P_{Y}^*(Y_t)}\middle|\mathcal{F}_{t-1}}
\end{align}
forms a martingale difference sequence and consequently, the sequence $Z'_t\! := \sum_{\tau=1}^t \tilde{Z}_{\tau}'$
forms a martingale. 
Analogously to \cite{wagner_new_2020}, we introduce the \textit{modified feedback strategy} as follows. For $t < n$ and $x^t \in \mathcal{X}^t$, let
\begin{align}
    \mathcal{X}_{x^t}^* &:= \{x: (x^t, x) \text{ is a prefix to some } x^n \text{ with } P_{x^n} \notin \mathcal{S}_{\gamma}^{n}\}, \\
    \mathcal{X}_{x^t}^T &:= \{x: (x^t, x) \text{ is a prefix to some } x^n \text{ with } P_{x^n} = T\}.
\end{align}

Now, for each feedback encoding function $F \in \mathcal{F}_n$, we define the modified feedback encoding function $\vec{f}_*$ as follows.
\begin{align}
    f_*^{t}(y^{t-1}) = \begin{cases}
        f^t(y^{t-1}), &\text{if } f^{t}(y^{t-1}) \in \mathcal{X}_{x^{t-1}}^*, \\
        x_0(x^{t-1}), &\text{otherwise},
    \end{cases}
\end{align}
where $x^{t-1}$ is the input history produced by the modified feedback strategy and $x_0(x^{t-1})$ denotes an arbitrary but fixed element from $\mathcal{X}_{x^{t-1}}^*$ or, if $\mathcal{X}_{x^{t-1}}^*$ is empty, $\mathcal{X}^*$. Define $\vec{f}_T$ analogously. Then the modified encoding distributions are
\begin{align}
    Q^*_F(\vec{g}) := \sum_{\vec{f} \in \mathcal{F}^n: \vec{f}_*=\vec{g}}Q_F(\vec{f})
\end{align}
and $Q^T_F(\vec{g})$ accordingly.
Furthermore, we define the auxiliary distribution $P_Y$ as follows, similar to \cite[eq. (87)]{wagner_new_2020}:
\begin{align}
    P_{Y^n}(y^n) = \frac{1}{2}\prod_{t=1}^n P_Y^*(y_t) + \frac{1}{2| \mathcal{S}_{\gamma}^{n}|} \sum_{T\in  \mathcal{S}_{\gamma}^{n}}\prod_{t=1}^n TW(y_t), \label{eq:idf-rand-converse-output-distribution}
\end{align}
where $P_Y^*$ is the capacity-achieving output distribution. If $\mathcal{S}_{\gamma}^{n} = \emptyset$, we set $P_{Y^n} = \prod_{t=1}^n P_Y^*(y_t)$. 
\begin{align}
    Q_FW^n(\mathcal{E}_{P_{Y^n}}^c) =& Q_F\times W^n(\mathcal{E}_{P_{Y^n}}^c \cap \psi_W(X^n) \leq \gamma) \nonumber\\
    &+ Q_F\times W^n(\mathcal{E}_{P_{Y^n}}^c \cap \psi_W(X^n) > \gamma) \nonumber \\
    =& Q_F\times W^n(\mathcal{E}_{P_{Y^n}}^c \cap \psi_W(X^n) \leq \gamma)\nonumber \\
    &+ \sum_{T \in \mathcal{S}_{\gamma}^n} Q_F\times W^n(\mathcal{E}_{P_{Y^n}}^c \cap \{P_{X^n} = T\}) \nonumber\\
    \leq& Q_F^*W^n(\mathcal{E}_{P_{Y^n}}^c)+ \sum_{T \in \mathcal{S}_{\gamma}^n} Q_F^T W^n(\mathcal{E}_{P_{Y^n}}^c). \label{eq:idf-rand-converse-prob-bound}
\end{align}
We start by analyzing the first probability in \eqref{eq:idf-rand-converse-prob-bound}. For brevity, define $\Pr_*[\cdot]$, $\expect[*]{\cdot}$, and $\Var_*[\cdot]$ with respect to the joint law generated by $Q^*_F$. Using \eqref{eq:idf-rand-converse-output-distribution}, we can bound
\begin{align}
    Q_F^*W^n(\mathcal{E}_{P_{Y^n}}^c) &= {\Pr}_*\left[\log \frac{1}{P_{Y}(Y^n)} > \log \rho_n  \right] \\
    &\leq {\Pr}_*\left[\sum_{t=1}^n \log \frac{1}{P_Y^*(Y_t)} > \log \rho_n - \log 2\right].
\end{align}

We will use the fact that
\begin{align}
    \expect[*]{\log\frac{1}{ P_Y^*(Y_t)}\middle|\mathcal{F}_{t-1}}
    &\leq H(P_Y^*),
\end{align}
Furthermore, we need bounds on the variance. Define
    \begin{align}
        \nu_{\mathsf{IDF}}^{x}(W) &:= \Var_{W(\cdot|x)}\left[\log \frac{1}{P_Y^*(Y)}\right],\\
        \nu_{\mathsf{IDF}}^{\max}(W) &:= \max_{x \in \mathcal{X}} \nu_{\mathsf{IDF}}^{x}(W).
    \end{align}
\begin{lemma}\label{lem:idf-rand-converse-lemma-closetypes}
    Assume that the dispersion fulfills $V_{\mathsf{IDF}}^{\min}(W) = V_{\mathsf{IDF}}^{\max}(W) =V_{\mathsf{IDF}}(W)>0$. 
    Then with $K_W' = \max\left\{2 \nu_{\mathsf{IDF}}^{\max}(W), \frac{8\nu_{\mathsf{IDF}}^{\max}(W)}{V_{\mathsf{IDF}}(W)}\right\}$  and $\gamma \in \left(0, \frac{V_{\mathsf{IDF}}(W)}{2K_W'}\right)$, we have 
    \begin{align}
       V_{\mathsf{IDF}}(W)-\gamma K_W' &\leq \frac{1}{n}\sum_{t=1}^n \expect[*]{(\tilde{Z}'_t)^{2}|\mathcal{F}_{t-1}} \nonumber\\
       &\leq V_{\mathsf{IDF}}(W) + \gamma K_W',
    \end{align}
    \begin{align}
        \left\|\frac{\sum_{t=1}^n \expect[*]{(\tilde{Z}'_t)^{2}|\mathcal{F}_{t-1}}}{\sum_{t=1}^n \expect[*]{(\tilde{Z}'_t)^{2}}}-1\right\|_{\infty} &\leq \gamma K_W'.\label{eq:idf-rand-proof-lemma-varbound-2}
    \end{align}
\end{lemma}
\begin{IEEEproof}
Analogously to the proof of Lemma \ref{lem:idf-det-converse-lemma-closetypes}, we obtain
\begin{align}
    \frac{1}{n}\sum_{t=1}^n\mathbb{E}_*\left[(\tilde{Z}_t')^2\mid \mathcal{F}_{t-1}\right] = \sum_{x\in\mathcal{X}} P_{X^n}(x) \nu_{\mathsf{IDF}}^{x}(W).
\end{align}
By construction of $\vec{f}_*$, there exists a $P \in \Phi_W^*$ such that
$d_{\mathrm{TV}}(P_{X^n}, P) = \psi_W(X^n) \leq \gamma.$
With $\sum_{x\in \mathcal{X}}P(x)\nu_{\mathsf{IDF}}^{x}(W) = V_{\mathsf{IDF}}(W),$
we obtain 
\begin{align}
    &\left|\sum_{x\in\mathcal{X}} P_{X^n}(x) \nu_{\mathsf{IDF}}^{x}(W)-V_{\mathsf{IDF}}(W)\right| \\
    &=\left|\sum_{x\in\mathcal{X}} \left(P_{X^n}(x) - P(x)\right) \nu_{\mathsf{IDF}}^{x}(W)\right| \\
    &\leq \nu_{\mathsf{IDF}}^{\max}(W)\sum_{x\in\mathcal{X}} \left|P_{X^n}(x) - P(x)\right| \\
    &= 2\nu_{\mathsf{IDF}}^{\max}(W)d_{\mathrm{TV}}(P_{X^n}, P) \\
    &\leq 2\gamma\nu_{\mathsf{IDF}}^{\max}(W).
\end{align}
The bound \eqref{eq:idf-rand-proof-lemma-varbound-2} is obtained analogously to the proof of Lemma \ref{lem:idf-det-converse-lemma-closetypes}.
\end{IEEEproof}
We abbreviate $H(P^*W) =: H$, define
\begin{align}
    V_{\gamma}'(u) := \begin{cases}
       V_{\mathsf{IDF}}(W)-\gamma K_W' &0 < u \leq \frac{1}{2} , \\
        V_{\mathsf{IDF}}(W)+\gamma K_W', &\frac{1}{2} < u < 1.
    \end{cases}
\end{align}
and let
\begin{align}
    \log \rho_n = nH - \sqrt{nV_{\gamma}'(\lambda_n)}Q^{-1} (\lambda_n) + \log 2,
\end{align}
where $\lambda_n$ is to be specified later.
Then we use Lemma \ref{lem:idf-rand-converse-lemma-closetypes} and the Martingale Berry-Esseen Theorem analogously to Section \ref{sec:idf-det-converse}, and find
\begin{align}
{\Pr}_*\left[\sum_{t=1}^n \log \frac{1}{P_Y^*(Y_t)} > \log \rho_n - \log 2\right] \leq 1-\lambda_n + \delta'_{1,n}, \label{eq:idf-rand-converse-closetypes-bound}
\end{align}
where $\delta'_{1,n} = L\left(\frac{\log n}{\sqrt{n}(V_{\mathsf{IDF}}(W)-\gamma K_W')^{3/2}}+\sqrt{\gamma K_W'}\right)$ is again an abbreviation for the Berry-Esseen ratio.

Now we consider the second term in \eqref{eq:idf-rand-converse-prob-bound}, i.e., the case that $P_{X^n} = T \in \mathcal{S}_{\gamma}^n$. Let $\Pr_T[\cdot]$ and $\expect[T]{\cdot}$ refer to the probabilies generated by the modified encoder $Q_F^T$. Using \eqref{eq:idf-rand-converse-output-distribution}, we find
\begin{align}
    Q_F^TW^n(\mathcal{E}_{P_{Y^n}}^c) &= {\Pr}_T\left[\log \frac{1}{P_{Y^n}(Y^n)} > \log \rho_n  \right] \\
    &\leq {\Pr}_T\left[\sum_{t=1}^n \log \frac{1}{TW(Y_t)} > \log \rho_n \!-\! \log 2|\mathcal{S}_{\gamma}^n|\right].
\end{align}

Now define
\begin{align}
    \mathcal{A}_{\gamma} := \left\{P \in \mathcal{P}(\mathcal{X}): d_{\mathrm{TV}}(P, \tilde{P}) \geq \gamma \;\forall \tilde{P} \in \Phi_W^*\right\}.
\end{align}
If $\mathcal{A}_{\gamma}$ is empty, then also $\mathcal{S}_{\gamma}^n$ is empty for every $n$ and we take $\delta_{2,n}'=0$. Otherwise, define
\begin{align}
    \Delta_{\gamma} &:= \min_{P\in \mathcal{A}_{\gamma}}[H(P*W)-H(PW)],\\
    K_T &:= H(P^*W) - H(TW) \geq \Delta_{\gamma} > 0.
\end{align}

Let $W_{\min} := \min_{x,y: W(y|x)>0} W(y|x)$ and define $Z_0^T = 0$ and
\begin{align}
    Z_t^T = \sum_{\tau = 1}^t \left(\log \frac{1}{TW(Y_{\tau})} - \mathbb{E}_T \left[\log \frac{1}{TW(Y_{\tau})} \middle| \mathcal{F}_{\tau-1}\right]\right),
\end{align}
$t > 0$. Note that 
\begin{align}
    \sum_{t=1}^n \expect[T]{\log \frac{1}{TW(Y_t)} \middle| \mathcal{F}_{t-1}}= nH(TW).
\end{align}
Then we have
\begin{align}
    &|Z_t^T - Z_{t-1}^T| \\
    &= \left|\log \frac{1}{TW(Y_t)}  - \mathbb{E}_T \left[\log \frac{1}{TW(Y_{t})} \middle| \mathcal{F}_{t-1}\right] \right| \\
    &\leq 2\log  \max_{y: TW(y) > 0} \frac{1}{TW(y)} \\
    &\leq 2\log  \max_{x, y: W(y|x) > 0} \frac{1}{W(y|x)}\max_{x: T(x) > 0} \frac{1}{T(x)}\\
    &\leq 2\log \frac{n}{W_{\min}} \label{eq:idf-rand-converse-martingale-difference-bound-intermediate}\\
    &\leq 3\log n,
\end{align}
where \eqref{eq:idf-rand-converse-martingale-difference-bound-intermediate} follows from $\min_{x: T(x) > 0}T(x) \geq \frac{1}{n}$ by the definition of the type and the last step hold for sufficiently large $n$.
We can now bound with the Azuma-Hoeffding inequality, similarly to the deterministic case in Section \ref{sec:idf-det-converse},
\begin{align}
    &{\Pr}_T\left[\sum_{t=1}^n \log \frac{1}{TW(Y_t)} > \log \rho_n \!-\! \log 2|\mathcal{S}_{\gamma}^n|\right] \\
    &\leq\exp\left(-\frac{nK_T^2}{32 \log^2 \frac{n}{W_{\min}}}\right)\\
    &\leq \exp\left(-\frac{nK'}{\log^2 n}\right),
\end{align}
where  we define $K':= \frac{\Delta_{\gamma}^2}{128} > 0$ and use $\log n/W_{\min} \leq 2\log n$ for $n$ sufficiently large. Note that $\log 2|\mathcal{S}_{\gamma}^n| \leq |\mathcal{X}|\log 2(n+1)$, and thus, for sufficiently large $n$, we can upper bound it by $\frac{nK_T}{2}$ together with the $\sqrt{n}$ term, similar to \eqref{eq:idf-det-converse-fartypes-nlarge}.
Summing over all possible type classes, we have 
\begin{align}
    &\sum_{T \in \mathcal{S}_{\gamma}^n} {\Pr}_T\left[\sum_{t=1}^n \log \frac{1}{TW(Y_t)} > \log \rho_n\!-\! \log 2|\mathcal{S}_{\gamma}^n|\right] \\
    &\leq (n+1)^{|\mathcal{X}|} \exp\left(-\frac{nK'}{\log^2 n}\right) =: \delta_{2,n}'. \label{eq:idf-rand-converse-fartypes-bound}
\end{align}
Inserting \eqref{eq:idf-rand-converse-closetypes-bound} and \eqref{eq:idf-rand-converse-fartypes-bound} into \eqref{eq:idf-rand-converse-prob-bound}, we obtain
\begin{align}
    Q_FW^n(\mathcal{E}_{P_{Y^n}}^c) &\leq 1-\lambda_n+\delta_{1,n}'+\delta_{2,n}'\\
    \Leftrightarrow Q_FW^n(\mathcal{E}_{P_{Y^n}}) &\geq \lambda_n-\delta_{1,n}'-\delta_{2,n}'.
\end{align}
The choice of $\lambda_n$ and the taking of the limits works exactly analogously to the proof of Theorem \ref{thm:second-order-idf} in Section \ref{sec:idf-det-converse}.

\section{Conclusion}\label{sec:conclusion}
We characterized the second-order capacity for randomized identification with feedback over any DMC and for deterministic identification with feedback for the class of simple-dispersion channels. Using a two-phase code construction and a new threshold-based receiver operation, we show the achievability of our result. The converse proofs rely on set counting arguments and martingale-based techniques to deal with the possible dependence between input symbols due to the feedback. We also show that for weakly symmetric channels, the second-order rate shows a suprising behavior: for deterministic identification with feedback, it is equal to the second-order coding rate of transmission, even though the capacities differ. For randomized identification with feedback, it equals zero, implying a fast convergence of the rates towards capacity. The latter result generalizes to all channels whose capacity-achieving output distribution is uniform, and it suggests the existence of codes operating very close to capacity even at smaller blocklengths. 

There are several open problems worth investigating. A natural extension of identification with noiseless feedback would be to consider noisy feedback \cite{ahlswede_new_1995}. However, the capacity of this setup is still unknown, which needs to be solved before second-order coding rates can be meaningfully derived. Another promising direction is joint identification and sensing \cite{labidi_joint_2024}, which introduces a channel estimation task on the transmitter side. For joint communication and sensing, bounds on the second-order coding rates have been found by Nikbakht et al. \cite{nikbakht_integrated_2024-2}, leaving a tight characterization as well as an extension to identification via channels still open.

\section*{Acknowledgement}
All content in this paper was derived and written by the human authors. The authors used the AI models GPT-6 and GPT-6.2 Astra for checking the proof, correcting errors and revising the manuscript.
All suggestions made by the AI were verified by the authors before using them in the manuscript.

\appendices
\appendix

\begin{lemma}[Weighted Bernoulli Chernoff Bound]\label{lem:weighted-chernoff-bernoulli}
    Let $0 < p < 1$ and $\rho > 0$. Let $Z_1, ..., Z_n$ be i.i.d. $\mathrm{Bern}(p)$ random variables, $0 \leq w_t \leq \frac{1}{\rho}$, $t \in [n]$ and $\sum_{t=1}^n w_t \leq 1$. Then for every $q \in (p, 1)$,
    \begin{align}
        \Pr\left[\sum_{t=1}^n w_tZ_t \geq q\right] \leq 2^{-\rho D(q\|p)}.
    \end{align}
\end{lemma}
\begin{IEEEproof}
    For any $s > 0$, we have
    \begin{align}
        \mathbb{E}\left[2^{s\rho\sum_{t=1}^n w_tZ_t}\right]
        &= \prod_{t=1}^n (1-p+p2^{s\rho w_t}) \\
        &\leq \prod_{t=1}^n (1-p+p2^{s})^{\rho w_t} \label{eq:bernoulli-chernoff-lemma-proof-ineq1}\\
        &\leq (1-p+p2^{s})^{\rho} \label{eq:bernoulli-chernoff-lemma-proof-ineq2}.
    \end{align}
    To see the first inequality, note that the function $g(u) = \log (1-p + p2^{su})$ is convex in $u$ and therefore, 
    \begin{align}
        g(u) &= g((1-u)\cdot 0 + u \cdot 1)\\
        &\leq (1-u)g(0) + ug(1) = ug(1)
    \end{align}
    Thus, 
    \begin{align}
        \log (1-p + p2^{su}) &\leq u \log (1-p + p2^{s}) \\
        \Leftrightarrow 1-p + p2^{su} &\leq (1-p + p2^{s})^{u}.
    \end{align}
    The inequality \eqref{eq:bernoulli-chernoff-lemma-proof-ineq1} follows by choosing $u = \rho w_t$ and \eqref{eq:bernoulli-chernoff-lemma-proof-ineq2} follows from $\sum_{t=1}^n w_t \leq 1$. Applying the Chernoff bound and using the Cramér transform for Bernoulli random variables as shown e.g. in \cite[Sec. 2.2]{boucheronConcentrationInequalitiesNonasymptotic2013}, we obtain
    \begin{align}
        \Pr\left[\sum_{t=1}^n w_tZ_t \geq q\right] &\leq \inf_{s > 0} 2^{-s\rho q}(1-p+p2^{s})^{\rho} \\
        &= 2^{-\rho D(q\|p)}.
    \end{align}
\end{IEEEproof}

\printbibliography

@article{ahlswede_identification_1989,
	title = {Identification via channels},
	volume = {35},
	doi = {10.1109/18.42172},
	pages = {15--29},
	number = {1},
	journaltitle = {{IEEE} Transactions on Information Theory},
	shortjournal = {{IEEE} Trans. Inf. Theory},
	author = {Ahlswede, R. and Dueck, G.},
	urldate = {2022-02-11},
	date = {1989-01},
}

@article{ahlswede_identification_1989-1,
	title = {Identification in the presence of feedback-a discovery of new capacity formulas},
	volume = {35},
	doi = {10.1109/18.42173},
	pages = {30--36},
	number = {1},
	journaltitle = {{IEEE} Transactions on Information Theory},
	shortjournal = {{IEEE} Trans. Inf. Theory},
	author = {Ahlswede, Rudolf and Dueck, Gunter},
	urldate = {2022-02-11},
	date = {1989-01},
}

@article{gunlu_code_2022,
	title = {Code Constructions and Bounds for Identification via Channels},
	volume = {70},
	doi = {10.1109/TCOMM.2021.3136864},
	pages = {1486--1496},
	number = {3},
	journaltitle = {{IEEE} Transactions on Communications},
	shortjournal = {{IEEE} Trans. Commun.},
	author = {Günlü, Onur and Kliewer, Jorg and Schaefer, Rafael F. and Sidorenko, Vladimir},
	urldate = {2022-02-14},
	date = {2022-03},
	langid = {english},
}

@article{salariseddigh_deterministic_2022,
	title = {Deterministic Identification Over Channels With Power Constraints},
	volume = {68},
	doi = {10.1109/TIT.2021.3122811},
	pages = {1--24},
	number = {1},
	journaltitle = {{IEEE} Transactions on Information Theory},
	shortjournal = {{IEEE} Trans. Inf. Theory},
	author = {Salariseddigh, Mohammad J. and Pereg, Uzi and Boche, Holger and Deppe, Christian},
	urldate = {2022-02-14},
	date = {2022-01},
	langid = {english},
}

@inproceedings{labidi_identification_2021,
	title = {Identification over the Gaussian Channel in the Presence of Feedback},
	doi = {10.1109/ISIT45174.2021.9517727},
	eventtitle = {2021 {IEEE} International Symposium on Information Theory ({ISIT})},
	pages = {278--283},
	booktitle = {Proc. {IEEE} Int. Symp. Inf. Theory ({ISIT})},
	author = {Labidi, Wafa and Boche, Holger and Deppe, Christian and Wiese, Moritz},
	urldate = {2022-02-14},
	date = {2021-07-12},
	langid = {english},
}

@article{derebeyoglu_performance_2020,
	title = {Performance Analysis of Identification Codes},
	volume = {22},
	doi = {10.3390/e22101067},
	pages = {1067},
	number = {10},
	journaltitle = {Entropy},
	shortjournal = {Entropy},
	author = {Derebeyoğlu, Sencer and Deppe, Christian and Ferrara, Roberto},
	urldate = {2022-02-14},
	date = {2020-09-23},
	langid = {english},
}

@article{verdu_explicit_1993,
	title = {Explicit construction of optimal constant-weight codes for identification via channels},
	volume = {39},
	doi = {10.1109/18.179339},
	pages = {30--36},
	number = {1},
	journaltitle = {{IEEE} Transactions on Information Theory},
	shortjournal = {{IEEE} Trans. Inf. Theory},
	author = {Verdú, S. and Wei, V.K.},
	urldate = {2022-02-14},
	date = {1993-01},
	langid = {english},
}

@inproceedings{labidi_secure_2020,
	title = {Secure Identification for Gaussian Channels},
	doi = {10.1109/ICASSP40776.2020.9054336},
	eventtitle = {{ICASSP} 2020 - 2020 {IEEE} International Conference on Acoustics, Speech and Signal Processing ({ICASSP})},
	pages = {2872--2876},
	booktitle = {Proc. {IEEE} Int. Conf. Acoust., Speech Signal Process. ({ICASSP})},
	author = {Labidi, Wafa and Deppe, Christian and Boche, Holger},
	urldate = {2022-02-15},
	date = {2020-05},
	langid = {english},
}

@article{ahlswede_new_1995,
	title = {New directions in the theory of identification via channels},
	volume = {41},
	doi = {10.1109/18.391249},
	pages = {1040--1050},
	number = {4},
	journaltitle = {{IEEE} Transactions on Information Theory},
	shortjournal = {{IEEE} Trans. Inf. Theory},
	author = {Ahlswede, R. and Zhang, Z.},
	urldate = {2022-02-15},
	date = {1995-07},
	langid = {english},
}

@article{molavianjazi_second-order_2015,
	title = {A Second-Order Achievable Rate Region for Gaussian Multi-Access Channels via a Central Limit Theorem for Functions},
	volume = {61},
	doi = {10.1109/TIT.2015.2492547},
	pages = {6719--6733},
	number = {12},
	journaltitle = {{IEEE} Transactions on Information Theory},
	shortjournal = {{IEEE} Trans. Inf. Theory},
	author = {{MolavianJazi}, Ebrahim and Laneman, J. Nicholas},
	urldate = {2022-02-02},
	date = {2015-12},
	langid = {english},
}

@article{tan_dispersions_2014,
	title = {On the Dispersions of Three Network Information Theory Problems},
	volume = {60},
	doi = {10.1109/TIT.2013.2291231},
	pages = {881--903},
	number = {2},
	journaltitle = {{IEEE} Transactions on Information Theory},
	shortjournal = {{IEEE} Trans. Inf. Theory},
	author = {Tan, Vincent Y. F. and Kosut, Oliver},
	urldate = {2022-02-08},
	date = {2014-02},
	langid = {english},
}

@article{polyanskiy_channel_2010,
	title = {Channel Coding Rate in the Finite Blocklength Regime},
	volume = {56},
	doi = {10.1109/TIT.2010.2043769},
	pages = {2307--2359},
	number = {5},
	journaltitle = {{IEEE} Transactions on Information Theory},
	shortjournal = {{IEEE} Trans. Inf. Theory},
	author = {Polyanskiy, Yury and Poor, H. Vincent and Verdú, Sergio},
	date = {2010-05},
	langid = {english},
}

@article{hayashi_information_2009,
	title = {Information Spectrum Approach to Second-Order Coding Rate in Channel Coding},
	volume = {55},
	doi = {10.1109/TIT.2009.2030478},
	pages = {4947--4966},
	number = {11},
	journaltitle = {{IEEE} Transactions on Information Theory},
	shortjournal = {{IEEE} Trans. Inf. Theory},
	author = {Hayashi, Masahito},
	date = {2009-11},
	langid = {english},
}

@inproceedings{strassen_asymptotische_1962,
	title = {Asymptotische Abschätzungen in Shannons Informationstheorie},
	pages = {689--723},
	booktitle = {Trans. 3rd Prague Conf. Inf. Theory},
	author = {Strassen, Volker},
	date = {1962},
}

@book{cover_elements_2006,
	location = {Hoboken, New Jersey},
	edition = {2},
	title = {Elements of Information Theory},
	pagetotal = {748},
	publisher = {John Wiley \& Sons},
	author = {Cover, Thomas M and Thomas, Joy A},
	date = {2006},
	langid = {english},
}

@book{gallager_information_1972,
	location = {Vienna},
	title = {Information Theory and Reliable Communication},
	isbn = {978-3-211-81145-0},
	url = {http://link.springer.com/10.1007/978-3-7091-2945-6},
	doi = {10.1007/978-3-7091-2945-6},
	publisher = {Springer Vienna},
	author = {Gallager, Robert},
	urldate = {2023-04-06},
	date = {1972},
}

@book{boucheronConcentrationInequalitiesNonasymptotic2013,
	location = {Oxford},
	edition = {1st ed},
	title = {Concentration inequalities: a nonasymptotic theory of independence},
	isbn = {978-0-19-953525-5},
	shorttitle = {Concentration inequalities},
	pagetotal = {481},
	publisher = {Oxford University Press},
	author = {Boucheron, Stéphane and Lugosi, Gábor and Massart, Pascal},
	date = {2013},
}

@article{bracher_identification_2017,
	title = {Identification via the Broadcast Channel},
	volume = {63},
	doi = {10.1109/TIT.2017.2674669},
	pages = {3480--3501},
	number = {6},
	journaltitle = {{IEEE} Transactions on Information Theory},
	shortjournal = {{IEEE} Trans. Inf. Theory},
	author = {Bracher, Annina and Lapidoth, Amos},
	date = {2017-06},
}

@article{watanabe_minimax_2022,
	title = {Minimax Converse for Identification via Channels},
	volume = {68},
	doi = {10.1109/TIT.2021.3120033},
	pages = {25--34},
	number = {1},
	journaltitle = {{IEEE} Transactions on Information Theory},
	shortjournal = {{IEEE} Trans. Inf. Theory},
	author = {Watanabe, Shun},
	date = {2022-01},
}

@book{ahlswede_identification_2021,
	location = {Cham},
	title = {Identification and Other Probabilistic Models: Rudolf Ahlswede’s Lectures on Information Theory 6},
	volume = {16},
	isbn = {978-3-030-65070-4},
	doi = {10.1007/978-3-030-65072-8},
	series = {Foundations in Signal Processing, Communications and Networking},
	shorttitle = {Identification and Other Probabilistic Models},
	publisher = {Springer International Publishing},
	author = {Ahlswede, Rudolf},
	editor = {Ahlswede, Alexander and Althöfer, Ingo and Deppe, Christian and Tamm, Ulrich},
	urldate = {2022-02-15},
	date = {2021},
	langid = {english},
}

@article{wagner_new_2020,
	title = {A New Method for Employing Feedback to Improve Coding Performance},
	volume = {66},
	doi = {10.1109/TIT.2020.2997385},
	pages = {6660--6681},
	number = {11},
	journaltitle = {{IEEE} Transactions on Information Theory},
	shortjournal = {{IEEE} Trans. Inf. Theory},
	author = {Wagner, Aaron B. and Shende, Nirmal V. and Altuğ, Yucel},
	urldate = {2023-11-22},
	date = {2020-11},
	langid = {english},
}

@article{fong_tight_2017,
	title = {A Tight Upper Bound on the Second-Order Coding Rate of the Parallel Gaussian Channel With Feedback},
	volume = {63},
	doi = {10.1109/TIT.2017.2740956},
	pages = {6474--6486},
	number = {10},
	journaltitle = {{IEEE} Transactions on Information Theory},
	shortjournal = {{IEEE} Trans. Inf. Theory},
	author = {Fong, Silas L. and Tan, Vincent Y. F.},
	urldate = {2023-11-22},
	date = {2017-10},
	langid = {english},
}

@inproceedings{altug_feedback_2014,
	title = {Feedback can improve the second-order coding performance in discrete memoryless channels},
	doi = {10.1109/ISIT.2014.6875256},
	eventtitle = {2014 {IEEE} International Symposium on Information Theory ({ISIT})},
	pages = {2361--2365},
	booktitle = {Proc. {IEEE} Int. Symp. Inf. Theory ({ISIT})},
	author = {Altuğ, Yucel and Wagner, Aaron B.},
	urldate = {2023-11-22},
	date = {2014-06},
	langid = {english},
}

@article{oohama_converse_2013-1,
	title = {Converse Coding Theorems for Identification via Channels},
	volume = {59},
	doi = {10.1109/TIT.2012.2219158},
	pages = {744--759},
	number = {2},
	journaltitle = {{IEEE} Transactions on Information Theory},
	shortjournal = {{IEEE} Trans. Inf. Theory},
	author = {Oohama, Yasutada},
	urldate = {2023-12-01},
	date = {2013-02},
	langid = {english},
}

@article{mross_gaussian_2024,
	title = {Gaussian Broadcast Channels With Heterogeneous Finite Blocklength Constraints: Inner and Outer Bounds},
	volume = {72},
	doi = {10.1109/TCOMM.2024.3354203},
	shorttitle = {Gaussian Broadcast Channels With Heterogeneous Finite Blocklength Constraints},
	pages = {2731--2745},
	number = {5},
	journaltitle = {{IEEE} Transactions on Communications},
	shortjournal = {{IEEE} Trans. Commun.},
	author = {Mross, Marcel A. and Lin, Pin-Hsun and Jorswieck, Eduard A.},
	urldate = {2024-05-28},
	date = {2024-05},
	langid = {english},
}

@article{yang_wiretap_2019,
	title = {Wiretap Channels: Nonasymptotic Fundamental Limits},
	volume = {65},
	doi = {10.1109/TIT.2019.2904500},
	shorttitle = {Wiretap Channels},
	pages = {4069--4093},
	number = {7},
	journaltitle = {{IEEE} Transactions on Information Theory},
	shortjournal = {{IEEE} Trans. Inf. Theory},
	author = {Yang, Wei and Schaefer, Rafael F. and Poor, H. Vincent},
	urldate = {2024-05-29},
	date = {2019-07},
	langid = {english},
}

@article{polyanskiy_feedback_2011,
	title = {Feedback in the Non-Asymptotic Regime},
	volume = {57},
	doi = {10.1109/TIT.2011.2158476},
	pages = {4903--4925},
	number = {8},
	journaltitle = {{IEEE} Transactions on Information Theory},
	shortjournal = {{IEEE} Trans. Inf. Theory},
	author = {Polyanskiy, Yury and Poor, H. Vincent and Verdú, Sergio},
	urldate = {2024-08-28},
	date = {2011-08},
}

@inproceedings{fong_asymptotic_2015,
	title = {Asymptotic expansions for the {AWGN} channel with feedback under a peak power constraint},
	doi = {10.1109/ISIT.2015.7282467},
	eventtitle = {2015 {IEEE} International Symposium on Information Theory ({ISIT})},
	pages = {311--315},
	booktitle = {Proc. {IEEE} Int. Symp. Inf. Theory ({ISIT})},
	author = {Fong, Silas L. and Tan, Vincent Y. F.},
	urldate = {2024-08-28},
	date = {2015-06},
}

@inproceedings{rosenberger_deterministic_2023,
	title = {Deterministic Identification Over Multiple-Access Channels},
	doi = {10.1109/ISIT54713.2023.10206481},
	eventtitle = {2023 {IEEE} International Symposium on Information Theory ({ISIT})},
	pages = {1699--1704},
	booktitle = {Proc. {IEEE} Int. Symp. Inf. Theory ({ISIT})},
	author = {Rosenberger, Johannes and Ibrahim, Abdalla and Deppe, Christian and Ferrara, Roberto},
	urldate = {2024-08-28},
	date = {2023-06},
}

@inproceedings{rosenberger_capacity_2023,
	title = {Capacity Bounds for Identification With Effective Secrecy},
	doi = {10.1109/ISIT54713.2023.10206589},
	eventtitle = {2023 {IEEE} International Symposium on Information Theory ({ISIT})},
	pages = {1202--1207},
	booktitle = {Proc. {IEEE} Int. Symp. Inf. Theory ({ISIT})},
	author = {Rosenberger, Johannes and Ibrahim, Abdalla and Bash, Boulat A. and Deppe, Christian and Ferrara, Roberto and Pereg, Uzi},
	urldate = {2024-08-29},
	date = {2023-06},
}

@inproceedings{jaja_identification_1985,
	title = {Identification is easier than decoding},
	doi = {10.1109/SFCS.1985.32},
	eventtitle = {26th Annual Symposium on Foundations of Computer Science ({SFCS} 1985)},
	pages = {43--50},
	booktitle = {Proc. 26th Annu. Symp. Found. Comput. Sci. ({SFCS})},
	author = {{JáJá}, Joseph},
	urldate = {2024-09-23},
	date = {1985-10},
}

@article{bolthausen_exact_1982,
	title = {Exact Convergence Rates in Some Martingale Central Limit Theorems},
	volume = {10},
	doi = {10.1214/aop/1176993776},
	pages = {672--688},
	number = {3},
	journaltitle = {The Annals of Probability},
	shortjournal = {Ann. Probab.},
	author = {Bolthausen, Erwin},
	urldate = {2024-11-08},
	date = {1982-08-01},
}

@incollection{cabrera_6g_2022,
	title = {6G and the Post‐Shannon Theory},
	doi = {10.1002/9781119765554.ch16},
	pages = {271--294},
	booktitle = {Shaping Future 6G Networks: Needs, Impacts, and Technologies},
	publisher = {{IEEE}},
	author = {Cabrera, Juan A. and Boche, Holger and Deppe, Christian and Schaefer, Rafael F. and Scheunert, Christian and Fitzek, Frank H. P.},
	urldate = {2025-03-06},
	date = {2022},
}

@inproceedings{liu_second-order_2024-1,
	title = {Second-Order Identification Capacity of {AWGN} Channels},
	doi = {10.1109/ISIT57864.2024.10619490},
	eventtitle = {2024 {IEEE} International Symposium on Information Theory ({ISIT})},
	pages = {309--314},
	booktitle = {Proc. {IEEE} Int. Symp. Inf. Theory ({ISIT})},
	author = {Liu, Zhicheng and Li, Yuan and Zhang, Huazi and Wang, Jun and Yan, Guiying and Ma, Zhiming},
	urldate = {2025-03-12},
	date = {2024-07},
}

@article{labidi_joint_2024,
	title = {Joint Identification and Sensing for Discrete Memoryless Channels},
	volume = {27},
	doi = {10.3390/e27010012},
	pages = {12},
	number = {1},
	journaltitle = {Entropy},
	shortjournal = {Entropy},
	author = {Labidi, Wafa and Zhao, Yaning and Deppe, Christian and Boche, Holger},
	urldate = {2025-03-19},
	date = {2024-12-27},
	langid = {english},
}

@article{salariseddigh_deterministic_2023,
	title = {Deterministic Identification for Molecular Communications Over the Poisson Channel},
	volume = {9},
	doi = {10.1109/TMBMC.2023.3324487},
	pages = {408--424},
	number = {4},
	journaltitle = {{IEEE} Transactions on Molecular, Biological, and Multi-Scale Communications},
	shortjournal = {{IEEE} Trans. Mol. Biol. Multi-Scale Commun.},
	author = {Salariseddigh, Mohammad Javad and Jamali, Vahid and Pereg, Uzi and Boche, Holger and Deppe, Christian and Schober, Robert},
	urldate = {2025-03-19},
	date = {2023-12},
}

@article{ahlswede_identification_1999,
	title = {Identification without randomization},
	volume = {45},
	doi = {10.1109/18.796419},
	pages = {2636--2642},
	number = {7},
	journaltitle = {{IEEE} Transactions on Information Theory},
	shortjournal = {{IEEE} Trans. Inf. Theory},
	author = {Ahlswede, R. and Cai, Ning},
	urldate = {2025-03-21},
	date = {1999-11},
}

@article{shannon_zero_1956,
	title = {The zero error capacity of a noisy channel},
	volume = {2},
	doi = {10.1109/TIT.1956.1056798},
	pages = {8--19},
	number = {3},
	journaltitle = {{IRE} Transactions on Information Theory},
	shortjournal = {{IRE} Trans. Inf. Theory},
	author = {Shannon, C.},
	urldate = {2025-03-21},
	date = {1956-09},
}

@incollection{bruche_converse_2025-1,
	location = {Cham},
	title = {Converse Techniques for Identification via Channels},
	isbn = {978-3-031-82014-4},
	doi = {10.1007/978-3-031-82014-4_13},
	pages = {317--361},
	booktitle = {Information Theory and Related Fields: Festschrift in Memory of Ning Cai},
	publisher = {Springer Nature Switzerland},
	author = {Brüche, Larissa and Mross, Marcel A. and Zhao, Yaning and Labidi, Wafa and Deppe, Christian and Jorswieck, Eduard A.},
	editor = {Deppe, Christian and Winter, Andreas and Yeung, Raymond W. and Boche, Holger and Althöfer, Ingo and Stoye, Jens and Tamm, Ulrich and Ezzine, Rami},
	date = {2025},
}

@inproceedings{mross_second-order_2025,
	location = {Sydney, Australia},
	title = {The Second-Order Coding Rate of Identification via Simple-Dispersion {DMCs} with Feedback},
	eventtitle = {2025 {IEEE} Inf. Theory Workshop ({ITW})},
	pages = {319--324},
	booktitle = {Proc. 2025 {IEEE} Inf. Theory Workshop ({ITW})},
	publisher = {{IEEE}},
	author = {Mross, Marcel A and Deppe, Christian and Jorswieck, Eduard A},
	date = {2025},
	langid = {english},
}

@misc{colomer_optimal_2026-1,
	title = {Optimal Codes for Deterministic Identification over Gaussian Channels: Closing the Capacity Gap},
	url = {http://arxiv.org/abs/2604.11782},
	doi = {10.48550/arXiv.2604.11782},
	shorttitle = {Optimal Codes for Deterministic Identification over Gaussian Channels},
	number = {{arXiv}:2604.11782},
	publisher = {{arXiv}},
	author = {Colomer, Pau and Deppe, Christian and Boche, Holger and Winter, Andreas},
	urldate = {2026-07-01},
	date = {2026-04-13},
	langid = {english},
	eprinttype = {arxiv},
	eprint = {2604.11782 [cs.IT]},
}

@article{von_lengerke_codes_2026,
	title = {Codes for Identification via Channels: Tutorial for Communications Generalists},
	volume = {28},
	rights = {https://ieeexplore.ieee.org/Xplorehelp/downloads/license-information/{IEEE}.html},
	issn = {1553-877X, 2373-745X},
	url = {https://ieeexplore.ieee.org/document/11036769/},
	doi = {10.1109/COMST.2025.3579815},
	shorttitle = {Codes for Identification via Channels},
	pages = {181--223},
	journaltitle = {{IEEE} Communications Surveys \& Tutorials},
	shortjournal = {{IEEE} Commun. Surv. Tutorials},
	author = {von Lengerke, Caspar and Cabrera, Juan A. and Reisslein, Martin and Fitzek, Frank H. P.},
	urldate = {2026-09-21},
	date = {2026},
	langid = {english},
}

@misc{labidi_identification_2026,
	title = {Identification Codes and Post-Shannon Communication: Theory, Architectures, and Emerging Applications},
	url = {http://arxiv.org/abs/2607.14666},
	doi = {10.48550/arXiv.2607.14666},
	shorttitle = {Identification Codes and Post-Shannon Communication},
	number = {{arXiv}:2607.14666},
	author = {Labidi, Wafa and Nilesh, Kumar and Rosenberger, Johannes and Cabrera, Juan and Boche, Holger and Deppe, Christian and Fitzek, Frank H. P. and Geitz, Marc},
	urldate = {2026-09-16},
	date = {2026-07-16},
	langid = {english},
	eprinttype = {arxiv},
	eprint = {2607.14666 [cs.IT]},
}

@inproceedings{nikbakht_integrated_2024-2,
	title = {Integrated Sensing and Communication in the Finite Blocklength Regime},
	issn = {2157-8117},
	url = {https://ieeexplore.ieee.org/document/10619685/},
	doi = {10.1109/ISIT57864.2024.10619685},
	eventtitle = {2024 {IEEE} International Symposium on Information Theory ({ISIT})},
	pages = {2790--2795},
	booktitle = {Proc. {IEEE} Int. Symp. Inf. Theory ({ISIT})},
	author = {Nikbakht, Homa and Wigger, Michéle and Shamai, Shlomo and Poor, H. Vincent},
	urldate = {2026-10-14},
	date = {2024-07},
}

@article{getu_survey_2024,
	title = {A Survey on Goal-Oriented Semantic Communication: Techniques, Challenges, and Future Directions},
	volume = {12},
	issn = {2169-3536},
	url = {https://ieeexplore.ieee.org/document/10479470/},
	doi = {10.1109/ACCESS.2024.3381967},
	shorttitle = {A Survey on Goal-Oriented Semantic Communication},
	pages = {51223--51274},
	journaltitle = {{IEEE} Access},
	author = {Getu, Tilahun M. and Kaddoum, Georges and Bennis, Mehdi},
	urldate = {2026-10-15},
	date = {2024},
}

@book{alon_probabilistic_2008,
	location = {Hoboken, N.J},
	edition = {3rd ed},
	title = {The probabilistic method},
	isbn = {978-0-470-17020-5},
	series = {Wiley-Interscience series in discrete mathematics and optimization},
	pagetotal = {352},
	publisher = {Wiley},
	author = {Alon, Noga and Spencer, Joel H.},
	date = {2008},
}

\end{document}